\documentclass[onecolumn,secnumarabic,amssymb, nobibnotes, aps, prd]{revtex4}
\usepackage{amsmath} \usepackage{amssymb}
\usepackage{graphicx} 
\usepackage{epstopdf}
\usepackage{amsmath}
\usepackage{amsmath,amssymb,amsthm,amsfonts,mathrsfs,bm,verbatim}
\usepackage{graphicx,subfigure}

\newcommand{\bea}{\begin{eqnarray}}
\newcommand{\eea}{\end{eqnarray}}

\begin{document}

\title{Thermodynamic geometric analysis of RN black holes under f(R) gravity }%

\author{Wen-Xiang Chen$^{1,a}$}
\affiliation{Department of Astronomy, School of Physics and Materials Science, GuangZhou University, Guangzhou 510006, China}
\email{1112019001@e.gzhu.edu.cn}
\author{Yao-Guang Zheng}
\affiliation{Department of Physics, College of Sciences, Northeastern University, Shenyang 110819, China}

\begin{abstract}
In this article, we explore the RN black hole under f(R) gravity and its thermodynamic properties. We begin by examining the small fluctuations around the equilibrium state and summarizing the expression for the modified thermodynamic entropy of this black hole. Additionally, we delve into the geometric thermodynamics (GTD) of black holes and investigate the suitability of the curvature scalar of the GTD method for the phase transition point of the black hole. Moreover, we investigate the effects of modified parameters on the thermodynamic behavior of black holes.Within the framework of $f(R)$ modified gravity theory, we discovered that several Reissner-Nordström (RN) black holes demonstrate thermodynamic properties resembling those of an ideal gas when the initial curvature scalar of the black hole remains constant. However, if the initial curvature scalar is non-constant and the cosmological constant term possesses a negative exponent, the Reissner-Nordström (RN) black holes could exhibit characteristics akin to those of a van der Waals gas.We separately list the general solutions for the case of non-negative powers and the special solutions for the case of negative powers. We observe that, under certain conditions, the phase transition analogous to the Van der Waals gas exists for charged black holes under f(R) gravity.

 \text { KEYwords: } f(R) \text { gravity; thermodynamic geometry; black hole. }

\end{abstract}

\maketitle

\section{Introduction}
Thermodynamic geometric analysis is a method of studying the thermodynamic properties of black holes based on their geometric properties and thermodynamic characteristics. Recently, scientists have started to conduct thermodynamic geometric analysis of Reissner-Nordström (RN) black holes under f(R) gravity.\cite{1,2,3,4,5,6}

RN black holes are a type of electrically charged black hole in which the gravitational attraction is balanced by the electromagnetic repulsion of the charged particles. The f(R) gravity theory is an alternative theory of gravity that modifies Einstein's general relativity by introducing a new function of the Ricci scalar curvature. This modification is believed to provide a more complete description of the behavior of gravity at both the cosmological and quantum scales.

The thermodynamic properties of black holes are usually expressed in terms of their entropy, temperature, and other thermodynamic variables. In the context of f(R) gravity, the thermodynamic properties of RN black holes have been studied using the geometric methods of Ruppeiner and Quevedo. These methods involve mapping the thermodynamic variables of the black hole to a thermodynamic surface, which is a geometrical representation of the thermodynamic state space.

The Ruppeiner geometry method is based on the idea of curvature, where the curvature of the thermodynamic surface is related to the fluctuations of the thermodynamic variables. The Quevedo geometry method, on the other hand, is based on the concept of information geometry, where the metric on the thermodynamic surface is related to the Fisher information metric.\cite{4,5,6,7,8}

The application of these geometric methods to the thermodynamic analysis of RN black holes under f(R) gravity has led to some interesting results. For example, it has been shown that the Ruppeiner curvature scalar for the RN black hole is negative, indicating the presence of repulsive interactions between the microscopic constituents of the black hole. This result is in contrast to the positive curvature scalar found in the case of non-rotating black holes in Einstein's general relativity.

Overall, the thermodynamic geometric analysis of RN black holes under f(R) gravity provides a new and interesting approach to studying the thermodynamic properties of black holes. This approach has the potential to shed light on the fundamental nature of gravity and its behavior at different scales, as well as to provide insights into the physics of black holes and the structure of the universe as a whole.

The study of black hole thermodynamic phase transition from the perspective of thermodynamic geometry is an important method in the study of phase transition\cite{4,5,6,7,8}. Therefore, we will construct three different thermodynamic geometries for this fine black hole in the parameter space based on the Hessian matrix, namely Weinhold geometry, Ruppeiner geometry, and free energy geometry. For these three different geometries, we will calculate their geometric standard curvatures separately. Furthermore, according to the specific form of the scalar curvature, the divergence behavior is discussed in depth, and the phase transition point and critical point of the black hole are compared to reveal the relationship between the phase transition point of the black hole and the singular curvature behavior.

In the discussed article, two scenarios are presented for the cosmological constant; interestingly, one isn't a constant term at all. This non-constant scenario is linked to f(R) gravity and either the scalar field or the electromagnetic tensor. Consequently, the cosmological constant we consider is intrinsically related to  $\pi$. However, Liu\cite{8} expanded this using thermodynamic geometry, in which the pressure term P will generate fractals, potentially leading to additional dimensions. Notably, this interpretation doesn't entirely align with f(R) gravity. In one f(R) gravity scenario, the geometric effects arise from the cumulative influence of all microscopic particles in the universe, rendering the cosmological constant a true constant. Yet, in another scenario, it's tied to the scalar field.

In the context of $f(R)$ modified gravitational theory, the Kerr-Newman black hole solution has been studied. Specifically, we have focused on the non-zero constant scalar curvature solution and investigated the metric tensor that satisfies the modified field equation.Furthermore, we have examined the thermodynamics of black holes and analyzed their local and global stability, without considering the presence of cosmological constants. This analysis has been conducted across various $f(R)$ models.By comparing our findings to those in general relativity, we have been able to highlight the main differences between the two theories. Our analysis has revealed a rich thermodynamic phenomenology that characterizes the $f(R)$ modified gravitational theory framework.

In this article, we investigate charged black holes in $f(R)$ gravity and derive modified thermodynamic entropy. We study the thermodynamic quantities and thermodynamic geometry of the black holes. We find that the Hawking temperature is inversely proportional to the event horizon radius. To evaluate the modifications to the static black hole entropy caused by thermal fluctuations in $f(R)$ gravity, we employ expressions for the Hawking temperature and the unmodified specific heat. We have observed that when the cosmological constant lacks negative power terms (i.e., when the initial curvature scalar is constant), the G-T diagram indicates flatness. However, when the cosmological constant includes negative power terms (i.e., when the initial curvature scalar is non-constant), the G-T diagram displays a comet-shaped structure. While it may be challenging to define these negative power terms precisely, we can still draw this conclusion. Within the framework of $f(R)$ modified gravity theory, we find that several Reissner-Nordström (RN) black holes exhibit thermodynamic characteristics similar to those of an ideal gas when the initial curvature scalar of the black hole is a constant. However, when the initial curvature scalar is non-constant and the cosmological constant term carries a negative power, the Reissner-Nordström (RN) black holes may display properties akin to a van der Waals gas.We delineate the general solutions for non-negative power cases alongside the special solutions for instances of negative powers. Notably, under specific conditions, the phase transition reminiscent of the Van der Waals gas can be observed in the context of charged black holes subject to f(R) gravity.

The article is structured as follows: Section 2 discusses the thermodynamic parameters in f(R) theory and thermodynamic entropy correction. Section 3 examines the thermodynamics of RN black holes in f(R) gravity background. Finally, Section 4 presents a summary and discussion.

\section{Thermodynamic parameters and thermodynamic entropy correction in f(R) theory}

In this section, we provide a summary of the expression for the thermodynamic entropy correction of black holes due to small fluctuations near equilibrium. To begin with, let us define the density of states with a fixed energy using natural units ($4 \pi=G=\hbar=c=1$) as follows \cite{9,10}:
\begin{equation}
\rho(E)=\frac{1}{2\pi i} \int_{c-i\infty}^{c+i\infty} e^{\mathcal{S}(\beta)} d\beta.
\end{equation}

The exact entropy $\mathcal{S}(\beta)=\log Z(\beta)+\beta E$ depends on the temperature $T(\beta^{-1})$ and is not just its equilibrium value. The exact entropy corresponds to the sum of the entropy of the thermodynamic system's subsystems, which is small enough to be considered in equilibrium. In order to study the form of the exact entropy, we solve the complex integral by using the method of steepest descent around the saddle point $\beta_{0}(=T_H^{-1})$ such that $\left.\frac{\partial \mathcal{S}(\beta)}{\partial \beta}\right|{\beta=\beta{0}}=0$. By performing a Taylor expansion of the exact entropy around the saddle point $\beta=\beta_{0}$, we obtain:
\begin{equation}
\mathcal{S}(\beta)={S_0} + \frac{1}{2}(\beta - \beta_0)^2 \left(\frac{\partial^2 \mathcal{S}(\beta)}{\partial \beta^2}\right){\beta=\beta_{0}} + (\text{higher order terms}).
\end{equation}

Thus, we can express the density of states as:
\begin{equation}
\rho(E) = \frac{e^{{S_0}}}{\sqrt{2\pi\left(\frac{\partial^2 \mathcal{S}(\beta)}{\partial \beta^2}\right){\beta=\beta_{0}}}}.
\end{equation}
where $c=\beta_{0}$ and $\left.\frac{\partial^2 \mathcal{S}(\beta)}{\partial \beta^2}\right|{\beta=\beta_{0}}>0$ are chosen.

Using the Wald relationship between the Noether charge of the differential homeomorphism and the entropy of a general spacetime with bifurcation surfaces, we introduce a method to obtain the effective family of higher-order derivatives from the entropy of black holes. Starting with the entropy, we analyze the derivation process of the action functional \cite{11,12,13}. This article claims that $4\pi$ equals 1.

\section{Thermodynamic of RN black holes in $f(R)$ gravity background}
In this section, we briefly review the main features of the four-dimensional charged AdS black hole corresponding in the $f(R)$ gravity background with a constant Ricci scalar curvature\cite{13,14}. It is possible for the cosmological constant to be negative and the initial curvature scalar to be positive. In this case, the negative cosmological constant would tend to decelerate the universe's expansion, while the positive initial curvature scalar would tend to accelerate it. The overall effect would depend on the relative magnitudes of the two quantities.So in this article, we set the cosmological constant as
\begin{equation}
{\Lambda_0} = -{\Lambda}.
\end{equation}There, $\Lambda_0$ is represented as the cosmological constant term, and $\Lambda$ is represented as a constant that is positively correlated with the initial curvature scalar.

The action is given by
\begin{equation}
S=\int_{\mathcal{M}} d^{4} x \sqrt{-g}\left[f(R)-F_{\mu \nu} F^{\mu \nu}\right].
\end{equation}

\subsection{Black Hole in the Form of f(R) Theory: $f(R)=R-\alpha R^{n}$}

In this example, we will examine the f(R) theory presented in references \cite{14}, which takes the form of
\begin{equation}
f(R)=R-\alpha R^{n}.
\end{equation}
In the $f(R)$ gravity background, the entropy of the RN black hole is expressed as follows:
\begin{equation}
S=\left(1-\alpha n R_{0}^{n-1}\right) \pi r_{+}^{2}=\frac{2-2 n}{2-n} \pi r_{+}^{2}.
\end{equation}
$\alpha$ and $n$ are constants such that $\alpha>0$ and $0<n<1$. The symbol $R$ represents the Ricci scalar curvature, and $f(R)$ is any arbitrary function of $R$. Moreover, $F_{\mu \nu}$ denotes the electromagnetic field tensor, which is given by $F_{\mu \nu}=\partial_{\mu} A_{\nu}-\partial_{\nu} A_{\mu}$, where $A_{\mu}$ is the electromagnetic potential. The equations of motion for the gravitational field $g_{\mu \nu}$ and the gauge field $A_{\mu}$, derived from the action, are as follows:
\begin{equation}
\begin{gathered}
R_{\mu \nu}\left[f^{\prime}(R)\right]-\frac{1}{2} g_{\mu \nu}[f(R)]+\left(g_{\mu \nu} \nabla^{2}-\nabla_{\mu} \nabla_{\nu}\right) f^{\prime}(R)=T_{\mu \nu} \\
\partial_{\mu}\left(\sqrt{-g} F^{\mu \nu}\right)=0.
\end{gathered}
\end{equation}
For a constant Ricci scalar curvature $R=R_{0}$, an analytical solution to the equations of motion has been obtained, where:
\begin{equation}
R_{\mu \nu}\left[f^{\prime}\left(R_{0}\right)\right]-\frac{g_{\mu \nu}}{4} R_{0}\left[f^{\prime}\left(R_{0}\right)\right]=T_{\mu \nu}.
\end{equation}
The $4$-dimensional gravity model for a charged static spherical black hole takes the form \cite{13}:
\begin{equation}
d s^{2}=-N(r) d t^{2}+\frac{d r^{2}}{N(r)}+r^{2}\left(d \theta^{2}+\sin ^{2} \theta d \phi^{2}\right)
\end{equation}
where the metric function $N(r)$ is given by:
\begin{equation}
N(r)=1-\frac{2 m}{r}+\frac{q^{2}}{b r^{2}}.
\end{equation}
Here, $b=f^{\prime}\left(R_{0}\right)$, and the parameters $m$ and $q$ are proportional to the black hole mass and charge, respectively \cite{14}:
\begin{equation}
M=m b, \quad Q=\frac{q}{\sqrt{b}}.
\end{equation}

In this context, we can obtain the electric potential $\Phi$ as follows:
\begin{equation}
\Phi=\frac{\sqrt{b} q}{r_{+}}.
\end{equation}
 $r_{+}$ corresponds to the largest root of the equation $N\left(r_{+}\right)=0$, which represents the black hole event horizon. At this point, we can also determine the Hawking temperature and entropy for this type of black hole \cite{13,14}, given by:
\begin{equation}
S=\pi r_{+}^{2} b.
\end{equation}
We can then calculate the Hawking temperature $\left(T_{H}=\frac{\partial M_{0}}{\partial {{S}_{0}}}\right)$ and the heat capacity $\left(C_{0}=T_{H} \frac{\partial {{S}_{0}}}{\partial T{H}}\right)$.

With the newly defined thermodynamic quantities, the first law of black hole thermodynamics in the extended phase space is expressed as:

\begin{equation}
dm = TdS + \Phi dQ + VdP.
\end{equation}

 m represents the mass, T is the temperature, S is the entropy, Q is the charge, $\Phi$ is the electrostatic potential,V is the thermodynamic volume, and P is the pressure.

Furthermore, the mass is interpreted as enthalpy in the extended phase space, and can be expressed as:
\begin{equation}
m = 2TS + \Phi Q - 2VP.
\end{equation}

The Hawking temperature under the conventional method is
\begin{equation}
T=(\frac{ m}{{r_+}^2}-\frac{2q^{2}}{b {r_+}^{3}})/(4\pi).
\end{equation}

By using the definition of Gibbs free energy, we can derive:
\begin{equation}
G = H - TS = m - TS = \frac{3Q^2}{4r_{+}} + \frac{r_{+}}{4},
\end{equation}
where $r_+$ is the radius of the event horizon.
\begin{figure}
  \centering
  \includegraphics[width=0.5\textwidth]{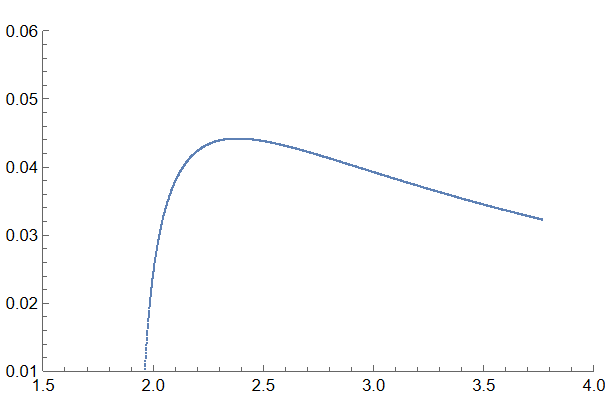}
  \caption{At this time, q=1, b=0.2, G is the horizontal axis, and T is the vertical axis}
  \label{1.png}
\end{figure}
\begin{figure}
  \centering
  \includegraphics[width=0.5\textwidth]{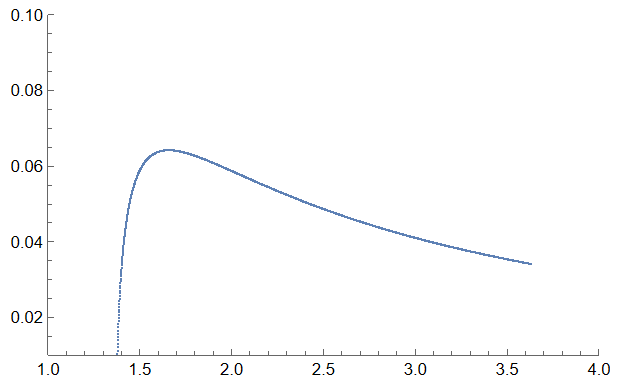}
  \caption{At this time, q=1, b=0.4, G is the horizontal axis, and T is the vertical axis}
  \label{1.1.png}
\end{figure}

Furthermore, we can express the metric as:
\begin{equation}
d s^{2}=-\frac{\partial^{2} S(r_{+},n)}{\partial X^{\alpha} \partial X^{\beta}} \Delta X^{\alpha} \Delta X^{\beta},
\end{equation}
where
\begin{equation}
g_{ij}=
\begin{pmatrix}
 \frac{4(-1+n) \pi}{-2+n} &   -\frac{4 \pi r_{+}}{(-2+n)^{2}} \\
 -\frac{4 \pi r_{+}}{(-2+n)^{2}}     &  \frac{4 \pi r_{+}^{2}}{(-2+n)^{3}}
\end{pmatrix}
\end{equation}
In addition, we have the ability to calculate the curvature scalar of the thermodynamic geometry, which can be expressed as follows:
\begin{equation}
R(S)=\frac{1}{8 \pi r_{+}^{2}}.
\end{equation}
If there is a naked singularity, the RN black hole will undergo a phase transition at that moment.

\subsection{Black hole in the form of f(R) theory: $f(R)=R-\lambda \exp (-\xi R)+\kappa R^{n}+\eta \ln (R)$}
The metric form is(The solution of RN black hole has a non-constant or constant Ricci curvature) \cite{15,16,17,18,19,20,21,22,23,24,25,26,27,28,29,30}, where
\begin{equation}
g(r)=k_1-\frac{2\Lambda_0}{(d-1)(d-2)} r^{2}-\frac{M}{r^{d-3}}+\frac{ Q^{2}}{r^{d-2}}.
\end{equation}
\begin{equation}
\begin{aligned}
&\lambda=\frac{R+\kappa R^{n}-\left(R+n \kappa R^{n}\right) \ln R}{(1+\xi R \ln R) e^{ -\xi R}}, \\
&\eta=-\frac{(1+\xi R) R+(n+\xi R) \kappa R^{n}}{1+\xi R \ln R},
\end{aligned}
\end{equation}where $\xi$ is a free parameter, and ${k_1}$ is a constant; it can take the values 1, -1, or 0. $M$ represents the mass of the black hole, while $\Lambda$ and $Q$ represent the cosmological constant and the charge of the black hole, respectively. Our objective is to determine the entropy $S$.

To calculate the entropy of the black hole, we first need to compute its Hawking temperature. The Hawking temperature can be obtained by evaluating the effect of surface gravity, derived from evaluating the black hole metric at the event horizon. In this case, the event horizon is located at $r=r_+$.

By solving the equation $g\left(r_{+}\right)=0$, we can obtain the value of $r_+$ that satisfies the equation.

\subsubsection{When d=4, ${k_1}$=1}
\begin{equation}
{M_1}=\frac{r_{+}}{2}\left(1+\frac{Q^2}{r_{+}^2}-\frac{\Lambda r_{+}^2}{3}\right).
\end{equation}Black hole pure mass is
\begin{equation}
M_p=\frac{r_{+}}{2}.
\end{equation}
The Hawking temperature
\begin{equation}
T=\frac{g^{\prime}\left(r_{+}\right)}{4 \pi}=\frac{1}{4 \pi}\left(\frac{1}{r_{+}}-\frac{Q^2}{r_{+}^3}-r_{+} \Lambda\right).
\end{equation}
Entropy is a measure of the amount of disorder or randomness in a system. It is typically denoted by the symbol S and can be calculated using the formula:
\begin{equation}
S=\int_0^{r_{+}} \frac{1}{T}\left(\frac{\partial{M_1}}{\partial r_{+}}\right) d r_{+}=\pi r_{+}^2.
\end{equation}
In the extended phase space, it is possible to consider the cosmological constant as a thermodynamic pressure and its conjugate quantity as thermodynamic volume. When the cosmological constant$\Lambda_0$ is negative
\begin{equation}
P=-\frac{\Lambda_0}{8 \pi} ,\quad V=\left(\frac{\partial {M_1}}{\partial P}\right)_{S, Q}.
\end{equation}
The thermodynamic volume $ V=\frac{4 \pi r_{+}^3}{3}$.
The Hawking temperature can be expressed as
\begin{equation}
T=\frac{1}{4 \pi}\left(\frac{1}{r_{+}}-\frac{Q^2}{r_{+}^3}+8 \pi P r_{+}\right).
\end{equation}
With the introduction of the newly defined thermodynamic quantities, the first law of black hole thermodynamics in the extended phase space can be expressed as follows:
\begin{equation}
d {M_1}=T d S+\Phi d Q+V d P.
\end{equation}
And 
\begin{equation}
{M_1}=2 T S+\Phi Q-2 V P.
\end{equation}
In the extended phase space, the mass is interpreted as enthalpy. So the
Gibbs free energy can be derived as

\begin{equation}
{G_1}=H-{T_1}S={M_1}-{T_1} S=\frac{3 Q^2}{4 r_{+}}+\frac{r_{+}}{4}-\frac{2 P \pi r_{+}^3}{3}
\end{equation}
\begin{figure}
  \centering
  \includegraphics[width=0.5\textwidth]{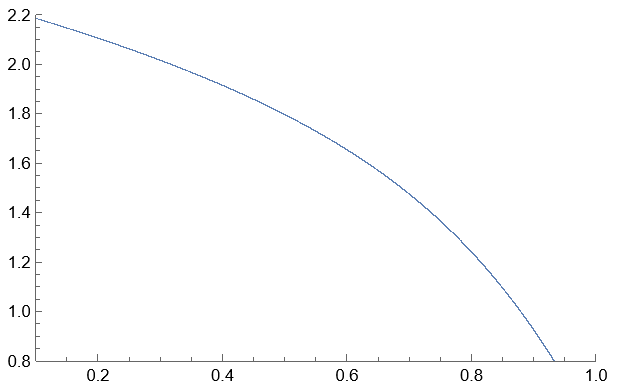}
  \caption{At this time, $Q$=1, P=0.4,$G_1$ is the horizontal axis, $T_1$ is the vertical axis}
  \label{2.png}
\end{figure}
\begin{figure}
  \centering
  \includegraphics[width=0.5\textwidth]{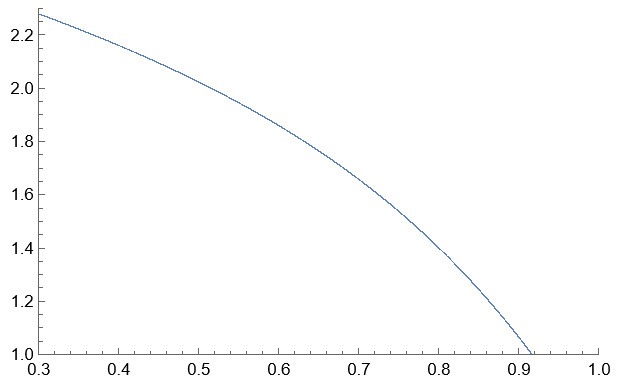}
  \caption{At this time, $Q$=1, P=0.5, $G_1$ is the horizontal axis, $T_1$ is the vertical axis}
  \label{2.1.png}
\end{figure}

To study the phase transition of free energy, considering the case where the cosmological constant is negative, and when the cosmological constant is positive, one can use R geometry to investigate its phase transition.

 We can express the metric as:\cite{29}
\begin{equation}
d s^{2}=-\frac{\partial^{2} S({M_1},Q)}{\partial X^{\alpha} \partial X^{\beta}} \Delta X^{\alpha} \Delta X^{\beta}.
\end{equation}
\begin{equation}
 r_+=\xi-\eta+\zeta, 
\end{equation}
where
\begin{equation}
\begin{aligned}
& \xi=\frac{1}{\sqrt{6}\Lambda_0}\left[1-\sqrt{1-12 Q^2 \Lambda_0^2} \cos \left(\frac{a}{3}-\frac{\pi}{3}\right)\right]^{1 / 2}, \\
& \eta=\frac{1}{\sqrt{6}\Lambda_0}\left[1-\sqrt{1-12 Q^2 \Lambda_0^2} \cos \left(\frac{a}{3}+\frac{\pi}{3}\right)\right]^{1 / 2}, \\
& \zeta=\frac{1}{\sqrt{6} \Lambda_0}\left[1+\sqrt{1-12 Q^2 \Lambda_0^2} \cos \left(\frac{a}{3}\right)\right]^{1 / 2}, \\
& a=\arccos \left[-\frac{1-18 \Lambda_0^2\left(3 M^2-2 Q^2\right)}{\left(1-12 Q^2 \Lambda_0^2\right)^{3 / 2}}\right] .
\end{aligned}
\end{equation}

Entropy is an important concept in thermodynamics that is often used to describe the degree of disorder or randomness in a system. The calculation of entropy involves the number of possible states of a system, and the number of states is related to the system's mass. Specifically, the larger the mass of a system, the greater the number of molecules or atoms it contains, and therefore, the greater the number of possible states. Thus, there exists a relationship between entropy and mass.

Regarding the relationship between entropy and electric charge, there is no direct connection between the two. When the second derivative of a function diverges, it may produce very large or very small changes near that point. Therefore, when the second derivative of entropy S with respect to mass M is infinite, the thermodynamic scalar curvature of entropy R(S) with respect to S is also infinite. We only list the denominator part:\cite{30,31,32,33,34,35}

Furthermore, $R$ can be rewritten in terms of number density $n=1 / 2 r_h$ as
\begin{equation}
R=\frac{1}{3 \pi} \frac{\left(n / n_c\right)^6-3\left(n / n_c\right)^4}{-\left(n / n_c\right)^4+6\left(n / n_c\right)^2+3\left(P / P_c\right)},
\end{equation}
where $Q=1$ and $\left(n_c, P_c\right)=(1 / 2 \sqrt{6}, 1 / 96 \pi)$.

\begin{equation}
g_{11}=\frac{d^2 S}{d {M_1}^2}\rightarrow 1/\left(\left(1-12 Q^2 \Lambda_0^2\right)^2\left(27{M_1}^4 \Lambda_0^2-{M_1}^2\left(1+36 Q^2 \Lambda_0^2\right)+\left(Q+4 Q^3 \Lambda_0^2\right)^2\right)\right).
\end{equation}

Additionally, we have the ability to calculate the curvature scalar of the thermodynamic geometry, which can be expressed as follows:We only list the denominator part of the curvature scalar R(S),
\begin{equation}
R(S)\rightarrow 1/\left(\left(1-12 Q^2 \Lambda_0^2\right)^2\left(27{M_1}^4 \Lambda_0^2-{M_1}^2\left(1+36 Q^2 \Lambda_0^2\right)+\left(Q+4 Q^3 \Lambda_0^2\right)^2\right)\right),
\end{equation}
when $\left(27 {M_1}^4 \Lambda_0^2-{M_1}^2\left(1+36 Q^2 \Lambda_0^2\right)+\left(Q+4 Q^3 \Lambda_0^2\right)^2\right)=0$ or $(1-12 Q^2 \Lambda_0^2)=0$, $R(S)$ diverges, indicating the presence of the phase transition.

\begin{equation}
P=\frac{T}{2 r_{+}}-\frac{1}{8 \pi r_{+}^2}+\frac{Q^2}{8 \pi r_{+}^4}, V=\frac{4 \pi r_{+}^3}{3}.
\end{equation}
The van der Waals equation provides a better description of the behavior of fluids above the critical temperature of gas-liquid phase transition compared to the ideal gas equation. It also provides a more reasonable description for liquids and low-pressure gases at temperatures slightly below the critical temperature. However, the volume (specific volume) changes with respect to changes in pressure, so the van der Waals equation is no longer applicable in such cases.

\begin{figure}
  \centering
  \includegraphics[width=0.7\textwidth]{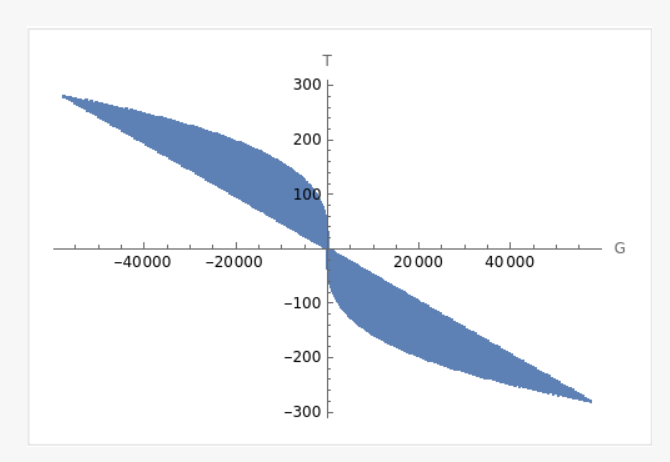}
  \caption{Here(section 3.2.1-FIG.5) is the plot of the function $G-T$ where $G_1$ is the Gibbs free energy, $T_1$ is the temperature, $Q=1, r_{+}$ranges from 1 to 14 , and $P$ is an algebraic number ranging from -10 to 10 .}
  \label{1.11.png}
\end{figure}From FIG.5, we can see that there are regions where the plot forms a "swallowtail" shape, which is indicative of a phase transition. This typically occurs when there are multiple solutions to the equations, and the system can transition between these solutions as the parameters are varied.

\begin{figure}
  \centering
  \includegraphics[width=0.7\textwidth]{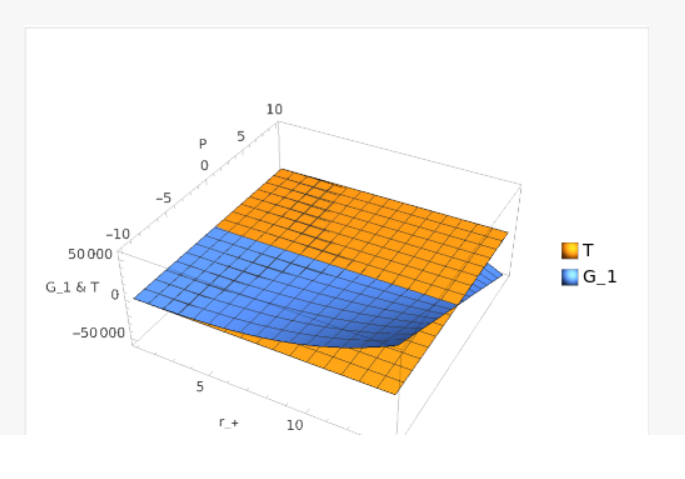}
  \caption{In Section 3.2.1, we present the plot of the function $G-T$. Here, $G_1$ represents the Gibbs free energy, $T_1$ denotes the temperature, $Q$ is set to $1, r_{+}$varies from 1 to 14 , and $P$ is an algebraic number that ranges between -10 and 10 }
  \label{1.111.png}
\end{figure}From the plot(FIG.6), it is evident that certain regions exhibit a 'swallowtail' shape, indicative of a phase transition. Such a pattern typically arises when multiple solutions exist for the equations, allowing the system to transition between these solutions as parameters change.

The significant difference between the ideal gas equation and the van der Waals equation lies in the fact that the ideal gas equation itself cannot predict phase transitions because its first-order phase transition point ($\partial p / \partial V=0$ ) has no solution, whereas the van der Waals equation does have phase transition points.

The phase transition of van der Waals gases can be described and predicted using the van der Waals equation, such as the van der Waals state equation. The van der Waals equation takes into account the intermolecular interactions, and by adjusting the van der Waals constants, the phase transition conditions, such as critical temperature and critical pressure, can be altered. In the van der Waals gas model, the phase transition points and processes can be determined and predicted through experiments or theoretical calculations. We observe that in the aforementioned process, the phase transition energy can be anticipated, which implies an approximation to the behavior of van der Waals gases.

\subsubsection{When d=4, ${k_1}$=-1}
We have the following equations:

\begin{equation}
{M_2}=\frac{r_{+}}{2}\left(-1+\frac{Q^2}{r_{+}^2}-\frac{\Lambda_0 r_{+}^2}{3}\right).
\end{equation}

The expression for the black hole pure mass is:

\begin{equation}
M=-\frac{r_{+}}{2}-\frac{\Lambda_0 r_{+}^2}{3}.
\end{equation}

The Hawking temperature is given by:

\begin{equation}
T=\frac{g^{\prime}\left(r_{+}\right)}{4 \pi}=\frac{1}{4 \pi}\left(\frac{-1}{r_{+}}-\frac{Q^2}{r_{+}^3}-r_{+} \Lambda_0\right).
\end{equation}

Entropy, denoted by the symbol $S$, is a measure of the amount of disorder or randomness in a system. It can be calculated using the formula:

\begin{equation}
S=\int_0^{r_{+}} \frac{1}{T}\left(\frac{\partial{M_2}}{\partial r_{+}}\right) d r_{+}=\pi r_{+}^2.
\end{equation}

In the extended phase space, the cosmological constant $\Lambda$ can be considered as a thermodynamic pressure, and its conjugate quantity is the thermodynamic volume $V$. When the cosmological constant $\Lambda$ is negative, we have:

\begin{equation}
P=-\frac{\Lambda_0}{8 \pi}, \quad V=\left(\frac{\partial {M_1}}{\partial P}\right)_{S, Q}.
\end{equation}

The thermodynamic volume $V$ is given by $V=\frac{4 \pi r_{+}^3}{3}$.

The Hawking temperature can be expressed as:

\begin{equation}
T_2=\frac{1}{4 \pi}\left(\frac{-1}{r_{+}}-\frac{Q^2}{r_{+}^3}+8 \pi P r_{+}\right).
\end{equation}

With the introduction of the newly defined thermodynamic quantities, the first law of black hole thermodynamics in the extended phase space can be expressed as follows:

\begin{equation}
d {M_2}=T d S+\Phi d Q+V d P.
\end{equation}

Furthermore,

\begin{equation}
{M_2}=2 T S+\Phi Q-2 V P.
\end{equation}

Gibbs free energy can be derived as
\begin{equation}
{G_2}=H-{T_1}S={M_2}-{T_1} S=\frac{3 Q^2}{4 r_{+}}-\frac{r_{+}}{4}-\frac{2 P \pi r_{+}^3}{3}
\end{equation}
\begin{figure}
  \centering
  \includegraphics[width=0.5\textwidth]{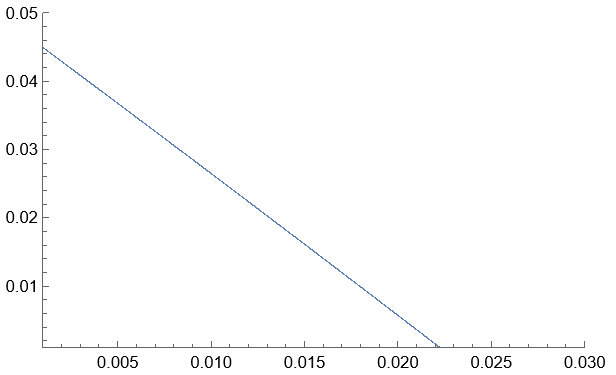}
  \caption{At this time, $Q$=1, P=0.4,$G_2$ is the horizontal axis, $T_2$ is the vertical axis}
  \label{4.png}
\end{figure}
\begin{figure}
  \centering
  \includegraphics[width=0.5\textwidth]{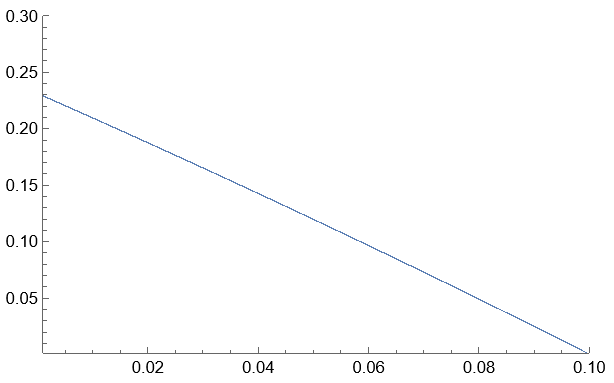}
  \caption{At this time, $Q$=1, P=0.5,$G_2$ is the horizontal axis, $T_2$ is the vertical axis}
  \label{5.png}
\end{figure}

Let g(r)=0, we get
\begin{equation}
\resizebox{\textwidth}{!}{$
\begin{aligned}
    r_+ &= \frac{1}{2} \sqrt{\frac{2}{\Lambda} + \frac{3 \cdot 2^{1/3} (1 + 4\Lambda Q^2)}{\Lambda \left(-54 + 243\Lambda {M_2}^2 + 648\Lambda Q^2 + \sqrt{-4(9 + 36\Lambda Q^2)^3 + (-54 + 243\Lambda M^2 + 648\Lambda Q^2)^2}\right)^{1/3}}} \\
    &+ \frac{1}{3 \cdot 2^{1/3} \Lambda} \left(-54 + 243\Lambda {M_2}^2 + 648\Lambda Q^2 + \sqrt{-4(9 + 36\Lambda Q^2)^3 + (-54 + 243\Lambda {M_2}^2 + 648\Lambda Q^2)^2}\right)^{1/3} \\
    &- \frac{1}{2} \sqrt{\frac{4}{\Lambda} - \frac{3 \cdot 2^{1/3} (1 + 4\Lambda Q^2)}{\Lambda \left(-54 + 243\Lambda {M_2}^2 + 648\Lambda Q^2 + \sqrt{-4(9 + 36\Lambda Q^2)^3 + (-54 + 243\Lambda {M_2}^2 + 648\Lambda Q^2)^2}\right)^{1/3}}} \\
    &- \frac{1}{3 \cdot 2^{1/3} \Lambda} \left(-54 + 243\Lambda {M_2}^2 + 648\Lambda Q^2 + \sqrt{-4(9 + 36\Lambda Q^2)^3 + (-54 + 243\Lambda {M_2}^2 + 648\Lambda Q^2)^2}\right)^{1/3} \\
    &+ \frac{6{M_2}}{\Lambda \sqrt{\frac{2}{\Lambda} + \frac{3 \cdot 2^{1/3} (1 + 4\Lambda Q^2)}{\Lambda \left(-54 + 243\Lambda {M_2}^2 + 648\Lambda Q^2 + \sqrt{-4(9 + 36\Lambda Q^2)^3 + (-54 + 243\Lambda {M_2}^2 + 648\Lambda Q^2)^2}\right)^{1/3}} + \frac{1}{3 \cdot 2^{1/3}\Lambda} \left(-54 + 243\Lambda {M_2}^2 + 648\Lambda Q^2 + \sqrt{-4(9 + 36\Lambda Q^2)^3 + (-54 + 243\Lambda {M_2}^2 + 648\Lambda Q^2)^2}\right)^{1/3}}}
\end{aligned}$}
\end{equation}

Regarding the relationship between entropy and electric charge, there is no direct connection between the two. When the second derivative of a function diverges, it can lead to significant changes near that point, either very large or very small. Therefore, if the second derivative of entropy (S) with respect to mass (M) is infinite, the thermodynamic scalar curvature of entropy (R(S)) with respect to S will also be infinite. Here, we will only focus on the denominator part:

\begin{equation}
g_{11}\rightarrow 1/(-54+243 \Lambda {M_2}^2+648 \Lambda Q^2+\sqrt{-4\left(9+36 \Lambda Q^2\right)^3+\left(-54+243\Lambda {M_2}^2+648 \Lambda Q^2\right)^2}).
\end{equation}

Furthermore, we can calculate the curvature scalar of the thermodynamic geometry, which can be expressed as follows. We will only list the denominator part of the curvature scalar R(S):

\begin{equation}
R(S)\rightarrow  1/(-54+243 \Lambda {M_2}^2+648 \Lambda Q^2+\sqrt{-4\left(9+36 \Lambda Q^2\right)^3+\left(-54+243\Lambda {M_2}^2+648 \Lambda Q^2\right)^2}).
\end{equation}

When $-54+243 \Lambda M^2+648 \Lambda Q^2+\sqrt{-4\left(9+36 \Lambda Q^2\right)^3+\left(-54+243\Lambda M^2+648 \Lambda Q^2\right)^2}=0$, $R(S)$ diverges, indicating the presence of the phase transitions.

\begin{figure}
  \centering
  \includegraphics[width=0.7\textwidth]{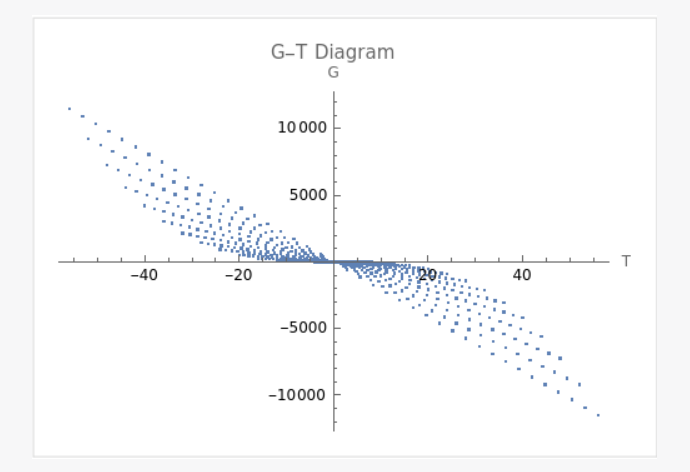}
  \caption{Here(Section 3.2.2) is the plot of the function $G-T$ where $G_2$ is the Gibbs free energy, $T_2$ is the temperature, $Q=1, r_{+}$ranges from 1 to 14 , and please note that this plot was generated with P values ranging from -2 to 2.}
  \label{2.11.png}
\end{figure}In FIG.9, we can see that under certain combinations of P and r values, the G-T graph exhibits a swallowtail shape. This typically occurs when the P value is close to 0 and the r value is close to 1. This is because under these conditions, the functional forms of G and T create complex interactions, resulting in a swallowtail shape in the plot.

\begin{figure}
  \centering
  \includegraphics[width=0.7\textwidth]{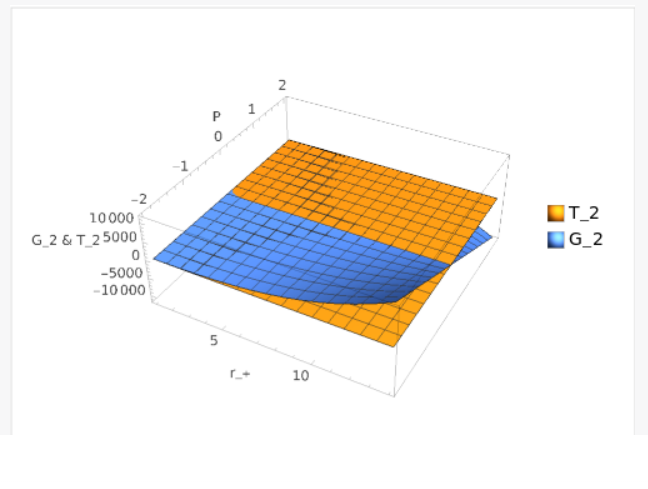}
  \caption{In Section 3.2.2, we present the plot of the function $G-T$. In this context, $G_2$ denotes the Gibbs free energy, $T_2$ represents the temperature, $Q$ is set to 1 , and $r_{+}$varies from 1 to 14 . It's important to note that this plot was generated using $P$ values ranging from -2 to 2. }
  \label{2.111.png}
\end{figure}In this plot(FIG.10), it's evident that the G-T graph assumes a swallowtail shape under specific combinations of P and r values. This phenomenon is particularly pronounced when the P value approaches 0 and the r value is near 1. Such a shape emerges due to the intricate interactions between the functional forms of G and T under these conditions.

\subsubsection{When d=4, ${k_1}$=0}
We have the following equations:

\begin{equation}
{M_3}=\frac{r_{+}}{2}\left(\frac{Q^2}{r_{+}^2}-\frac{\Lambda_0 r_{+}^2}{3}\right).
\end{equation}

The expression for the black hole pure mass is given :
\begin{equation}
M=-\frac{\Lambda_0 r_{+}^2}{3}.
\end{equation}

The Hawking temperature is given:
\begin{equation}
T=\frac{g^{\prime}\left(r_{+}\right)}{4 \pi}=\frac{1}{4 \pi}\left(-\frac{Q^2}{r_{+}^3}-r_{+} \Lambda_0\right).
\end{equation}

Entropy, denoted by the symbol $S$, is a measure of the amount of disorder or randomness in a system. It can be calculated using the formula given :
\begin{equation}
S=\int_0^{r_{+}} \frac{1}{T}\left(\frac{\partial{M_3}}{\partial r_{+}}\right) d r_{+}=\pi r_{+}^2.
\end{equation}

In the extended phase space, the cosmological constant $\Lambda$ can be considered as a thermodynamic pressure, and its conjugate quantity is the thermodynamic volume $V$. When the cosmological constant $\Lambda$ is negative, we have:

\begin{equation}
P=-\frac{\Lambda_0}{8 \pi}, \quad V=\left(\frac{\partial {M_1}}{\partial P}\right)_{S, Q}.
\end{equation}

The thermodynamic volume $V$ is given by $V=\frac{4 \pi r_{+}^3}{3}$.

The Hawking temperature can be expressed as :
\begin{equation}
T_3=\frac{1}{4 \pi}\left(-\frac{Q^2}{r_{+}^3}+8 \pi P r_{+}\right).
\end{equation}

With the introduction of the newly defined thermodynamic quantities, the first law of black hole thermodynamics in the extended phase space can be expressed as follows:

\begin{equation}
d {M_3}=T d S+\Phi d Q+V d P.
\end{equation}

Furthermore, the Equation  states:
\begin{equation}
{M_3}=2 T S+\Phi Q-2 V P.
\end{equation}

Finally, it shows the expression for ${G_3}$:
\begin{equation}
{G_3}=H-{T_3}S={M_3}-{T_3} S=\frac{3 Q^2}{4 r_{+}}-\frac{2 P \pi r_{+}^3}{3}.
\end{equation}
\begin{figure}
  \centering
  \includegraphics[width=0.5\textwidth]{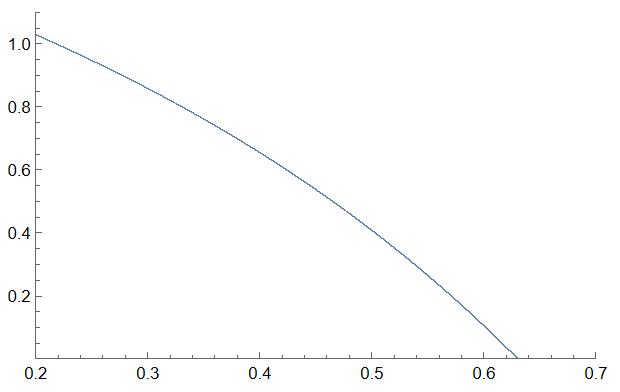}
  \caption{At this time, Q=1, P=0.4,$G_3$ is the horizontal axis, $T_3$ is the vertical axis}
  \label{6.png}
\end{figure}
\begin{figure}
  \centering
  \includegraphics[width=0.5\textwidth]{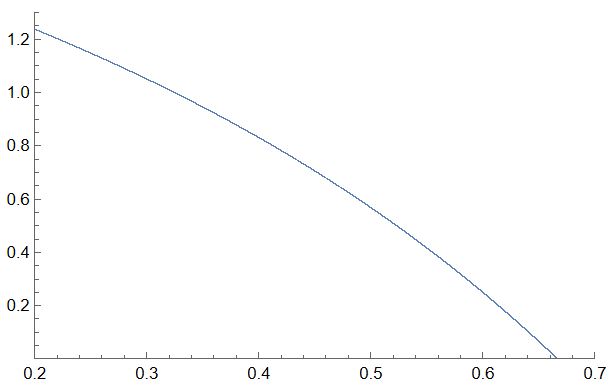}
  \caption{At this time, Q=1, P=0.5,$G_3$ is the horizontal axis, $T_3$ is the vertical axis}
  \label{7.png}
\end{figure}

When g(r)=0, we get
\begin{equation}
\resizebox{\textwidth}{!}{$
\begin{aligned}
\left\{r_+ = \frac{1}{2}\right. & \sqrt{\frac{4 \times 2^{1 / 3} Q^2}{\left(9\Lambda {M_3}^2+\sqrt{81 \Lambda^2 {M_3}^4-256 \Lambda^3 Q^6}\right)^{1 / 3}}+\frac{\left(9 \Lambda {M_3}^2+\sqrt{81 \Lambda^2 {M_3}^4-256 \Lambda^3 Q^6}\right)^{1 / 3}}{2^{1 / 3} \Lambda}}- \\
& \frac{1}{2} \sqrt{-\frac{4 \times 2^{1 / 3} Q^2}{\left(9 \Lambda {M_3}^2+\sqrt{81 \Lambda^2 {M_3}^4-256 \Lambda^3 Q^6}\right)^{1 / 3}}-\frac{\left(9 \Lambda {M_3}^2+\sqrt{81 \Lambda^2 {M_3}^4-256 \Lambda^3 Q^6}\right)^{1 / 3}}{2^{1 / 3} \Lambda}+\frac{6 M}{\left.\Lambda \sqrt{\frac{4 \cdot 2^{1 / 3} Q^2}{\left(9 \Lambda{M_3}^2+\sqrt{81 \Lambda^2 {M_3}^4-256 \Lambda^3 Q^6}\right)^{1 / 3}}+\frac{\left(9 \Lambda {M_3}^2+\sqrt{81 \Lambda^2 {M_3}^4-256 \Lambda^3 Q^6}\right)^{1 / 3}}{2^{1 / 3} \Lambda}}\right\}}},
\end{aligned}$}
\end{equation}

Regarding the relationship between entropy and electric charge, there is no direct connection between the two. When the second derivative of a function diverges, it can lead to significant changes near that point, either very large or very small. Therefore, if the second derivative of entropy (S) with respect to mass (M) is infinite, the thermodynamic scalar curvature of entropy (R(S)) with respect to S will also be infinite. In this discussion, we will focus on the denominator part:

\begin{equation}
g_{11}\rightarrow \frac{1}{9 {M_3}^2 \Lambda+\sqrt{81 {M_3}^4 \Lambda^2-256 Q^6 \Lambda^3}}.
\end{equation}

Furthermore, we can calculate the curvature scalar of the thermodynamic geometry, which can be expressed as follows. We will only list the denominator part of the curvature scalar R(S):

\begin{equation}
R(S)\rightarrow \frac{1}{9 {M_3}^2 \Lambda+\sqrt{81 {M_3}^4 \Lambda^2-256 Q^6 \Lambda^3}}.
\end{equation}
When $9 M^2 \Lambda+\sqrt{81 M^4 \Lambda^2-256 Q^6 \Lambda^3}=0$, $R(S)$ diverges, indicating the presence of phase transitions.
\begin{figure}
  \centering
  \includegraphics[width=0.7\textwidth]{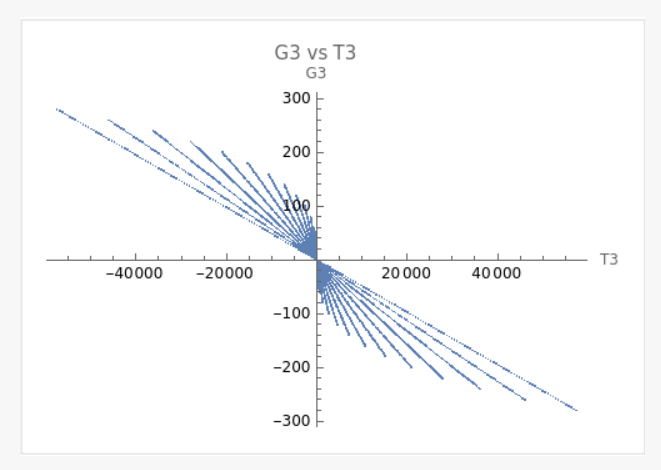}
  \caption{Here(section 3.2.3) is the plot of the function $G-T$ where $G2$ is the Gibbs free energy, $T2$ is the temperature, $Q=1, r_{+}$ranges from 1 to 14 , and please note that this plot was generated with P values ranging from -2 to 2.}
  \label{3.11.png}
\end{figure}
From the plot(FIG.13), it can be seen that the ``swallowtail" shape  appears in certain regions. The exact conditions for this shape to appear would require a more detailed analysis.

\begin{figure}
  \centering
  \includegraphics[width=0.7\textwidth]{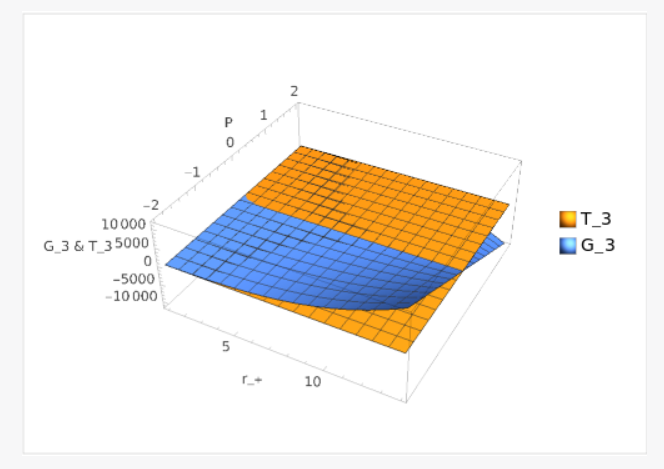}
  \caption{In Section 3.2.3, we present the plot of the function $G-T$. Here, $G_3$ represents the Gibbs free energy, $T_3$ denotes the temperature, $Q$ is set to 1 , and $r_{+}$varies from 1 to 14 . It's important to note that this plot was generated using $P$ values ranging from -2 to 2 .}
  \label{3.111.png}
\end{figure}
From FIG.14, the distinctive 'swallowtail' shape is evident in specific regions. A more detailed analysis would be necessary to pinpoint the exact conditions under which this shape emerges.

\subsubsection{When d=3, ${k_1}$=1}
We have the following equations:

\begin{equation}
g(r)=-\Lambda_0 r^{2}-M +\frac{Q^{2}}{r}+1.
\end{equation}

At this point, the Ricci scalar is obtained by setting $g(r)=0$:

\begin{equation}
\begin{aligned}
& \left\{\left\{r_+ = \frac{2^{1 / 3}(-3 \Lambda+3 M \Lambda)}{3 \Lambda\left(-27 Q^2 \Lambda^2+\sqrt{729 Q^4 \Lambda^4+4(-3 \Lambda+3 M \Lambda)^3}\right)^{1 / 3}}-\right.\right. \\
& \left.\frac{\left(-27 Q^2 \Lambda^2+\sqrt{729 Q^4 \Lambda^4+4(-3 \Lambda+3 M \Lambda)^3}\right)^{1 / 3}}{3 \times 2^{1 / 3} \Lambda}\right\}, \\
&
\end{aligned}
\end{equation}
where $r_+$ is the event horizon radius and the unique Killing horizon radius.

When g(r)=0.we get 
\begin{equation}
M=M_4=-\Lambda_0 r^{2} +\frac{Q^{2}}{r}+1.
\end{equation}

The calculation of the Hawking temperature using the conventional method is as follows:
\begin{equation}
T_4=-\frac{\Lambda_0 r_{+}}{2 \pi}-\frac{2Q^{2}}{ r_{+}^{2}}.
\end{equation}

The Gibbs free energy can be derived as:

\begin{equation}
G_4=\Lambda_0 r_{+}^{2}.
\end{equation}

The entropy for this BTZ-f(R) black hole solution is:

\begin{equation}
S=4 \pi r_{+}.
\end{equation}

\begin{figure}
\centering
\includegraphics[width=0.5\textwidth]{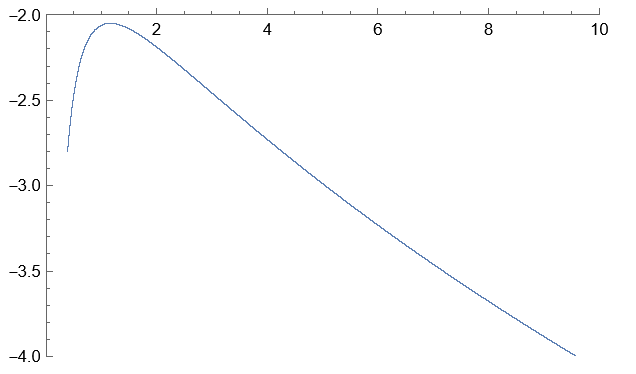}
\caption{At this time, Q=1,$\Lambda_0$=0.4. $G_4$ is the horizontal axis, and $T_4$ is the vertical axis.}
\label{10.png}
\end{figure}

\begin{figure}
\centering
\includegraphics[width=0.5\textwidth]{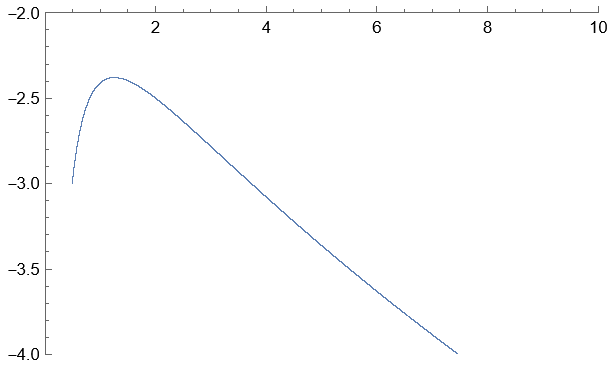}
\caption{At this time, Q=1, $\Lambda_0$=0.5. $G_4$ is the horizontal axis, and $T_4$ is the vertical axis.}
\label{11.png}
\end{figure}

We can express the metric as:

\begin{equation}
d s^{2}=-\frac{\partial^{2} S({M_4},\Lambda)}{\partial X^{\alpha} \partial X^{\beta}} \Delta X^{\alpha} \Delta X^{\beta},
\end{equation}

where
\begin{equation}\resizebox{\textwidth}{!}{$
\begin{aligned}
\frac{d^2{S}}{d {M_4}^2}=\left(48 \times 2^{1 / 3} \Lambda(-3 \Lambda+3 M \Lambda)^5\right) /\left(\left(729 Q^4 \Lambda^4+4(-3 \Lambda+3 M \Lambda)^3\right)\left(-27 Q^2 \Lambda^2+\sqrt{729 Q^4 \Lambda^4+4(-3 \Lambda+3 M \Lambda)^3}\right)^{7 / 3}\right)+\\
\left(12 \times 2^{2 / 3} \Lambda(-3 \Lambda+3 M \Lambda)^4\right) /\left(\left(729 Q^4 \Lambda^4+4(-3 \Lambda+3 M \Lambda)^3\right)\left(-27 Q^2 \Lambda^2+\sqrt{729 Q^4 \Lambda^4+4(-3 \Lambda+3 M \Lambda)^3}\right)^{5 / 3}\right)+\\
\left(36 \times 2^{1 / 3} \Lambda(-3 \Lambda+3 M \Lambda)^5\right) /\left(\left(729 Q^4 \Lambda^4+4(-3 \Lambda+3 M \Lambda)^3\right)^{3 / 2}\left(-27 Q^2 \Lambda^2+\sqrt{729 Q^4 \Lambda^4+4(-3 \Lambda+3 M \Lambda)^3}\right)^{4 / 3}\right)-\\
\left(24 \times 2^{1 / 3} \Lambda(-3 \Lambda+3 M \Lambda)^2\right) /\left(\sqrt{729 Q^4 \Lambda^4+4(-3 \Lambda+3 M \Lambda)^3}\left(-27 Q^2 \Lambda^2+\sqrt{729 Q^4 \Lambda^4+4(-3 \Lambda+3 M \Lambda)^3}\right)^{4 / 3}\right)+\\
\left(18 \times 2^{2 / 3} \Lambda(-3 \Lambda+3 M \Lambda)^4\right) /\left(\left(729 Q^4 \Lambda^4+4(-3 \Lambda+3 M \Lambda)^3\right)^{3 / 2}\left(-27 Q^2 \Lambda^2+\sqrt{729 Q^4 \Lambda^4+4(-3 \Lambda+3 M \Lambda)^3}\right)^{2 / 3}\right)-\\
\left(6 \times 2^{2 / 3} \Lambda(-3 \Lambda+3 M \Lambda)\right) /\left(\sqrt{729 Q^4 \Lambda^4+4(-3 \Lambda+3 M \Lambda)^3}\left(-27 Q^2 \Lambda^2+\sqrt{729 Q^4 \Lambda^4+4(-3 \Lambda+3 M \Lambda)^3}\right)^{2 / 3}\right),
\end{aligned}$}
\end{equation}
\begin{equation}
\resizebox{1\hsize}{!}{$
\begin{aligned}
\frac{d^2 {S}}{d{M_4}d \Lambda} = \frac{2\ 2^{2/3} (3 M \Lambda -3 \Lambda )^2 \left(\frac{2916 \Lambda ^3 Q^4+12 (3 M-3) (3 M \Lambda -3 \Lambda )^2}{2 \sqrt{729 Q^4 \Lambda ^4+4 (3 M \Lambda -3 \Lambda )^3}}-54 Q^2 \Lambda \right)}{3 \sqrt{729 Q^4 \Lambda ^4+4 (3 M \Lambda -3 \Lambda )^3} \left(\sqrt{729 Q^4 \Lambda ^4+4 (3 M \Lambda -3 \Lambda )^3}-27 Q^2 \Lambda ^2\right)^{5/3}}+\\
\frac{8 \sqrt[3]{2} (3 M \Lambda -3 \Lambda )^3 \left(\frac{2916 \Lambda ^3 Q^4+12 (3 M-3) (3 M \Lambda -3 \Lambda )^2}{2 \sqrt{729 Q^4 \Lambda ^4+4 (3 M \Lambda -3 \Lambda )^3}}-54 Q^2 \Lambda \right)}{3 \sqrt{729 Q^4 \Lambda ^4+4 (3 M \Lambda -3 \Lambda )^3} \left(\sqrt{729 Q^4 \Lambda ^4+4 (3 M \Lambda -3 \Lambda )^3}-27 Q^2 \Lambda ^2\right)^{7/3}}-\\
\frac{\frac{36 (3 M \Lambda -3 \Lambda )^2+72 (3 M-3) \Lambda  (3 M \Lambda -3 \Lambda )}{2 \sqrt{729 Q^4 \Lambda ^4+4 (3 M \Lambda -3 \Lambda )^3}}-\frac{9 \Lambda  (3 M \Lambda -3 \Lambda )^2 \left(2916 \Lambda ^3 Q^4+12 (3 M-3) (3 M \Lambda -3 \Lambda )^2\right)}{\left(729 Q^4 \Lambda ^4+4 (3 M \Lambda -3 \Lambda )^3\right)^{3/2}}}{9 \sqrt[3]{2} \Lambda  \left(\sqrt{729 Q^4 \Lambda ^4+4 (3 M \Lambda -3 \Lambda )^3}-27 Q^2 \Lambda ^2\right)^{2/3}}+\frac{2^{2/3} (3 M \Lambda -3 \Lambda )^2}{\Lambda  \sqrt{729 Q^4 \Lambda ^4+4 (3 M \Lambda -3 \Lambda )^3} \left(\sqrt{729 Q^4 \Lambda ^4+4 (3 M \Lambda -3 \Lambda )^3}-27 Q^2 \Lambda ^2\right)^{2/3}}-\\
\frac{\sqrt[3]{2} \left(\frac{2916 \Lambda ^3 Q^4+12 (3 M-3) (3 M \Lambda -3 \Lambda )^2}{2 \sqrt{729 Q^4 \Lambda ^4+4 (3 M \Lambda -3 \Lambda )^3}}-54 Q^2 \Lambda \right)}{3 \left(\sqrt{729 Q^4 \Lambda ^4+4 (3 M \Lambda -3 \Lambda )^3}-27 Q^2 \Lambda ^2\right)^{4/3}}-\frac{\sqrt[3]{2} (3 M \Lambda -3 \Lambda ) \left(\frac{36 (3 M \Lambda -3 \Lambda )^2+72 (3 M-3) \Lambda  (3 M \Lambda -3 \Lambda )}{2 \sqrt{729 Q^4 \Lambda ^4+4 (3 M \Lambda -3 \Lambda )^3}}-
\frac{9 \Lambda  (3 M \Lambda -3 \Lambda )^2 \left(2916 \Lambda ^3 Q^4+12 (3 M-3) (3 M \Lambda -3 \Lambda )^2\right)}{\left(729 Q^4 \Lambda ^4+4 (3 M \Lambda -3 \Lambda )^3\right)^{3/2}}\right)}{9 \Lambda  \left(\sqrt{729 Q^4 \Lambda ^4+4 (3 M \Lambda -3 \Lambda )^3}-27 Q^2 \Lambda ^2\right)^{4/3}}-\\
\frac{2 \sqrt[3]{2} (3 M-3) (3 M \Lambda -3 \Lambda )^2}{\sqrt{729 Q^4 \Lambda ^4+4 (3 M \Lambda -3 \Lambda )^3} \left(\sqrt{729 Q^4 \Lambda ^4+4 (3 M \Lambda -3 \Lambda )^3}-27 Q^2 \Lambda ^2\right)^{4/3}}+\frac{2 \sqrt[3]{2} (3 M \Lambda -3 \Lambda )^3}{\Lambda  \sqrt{729 Q^4 \Lambda ^4+4 (3 M \Lambda -3 \Lambda )^3} \left(\sqrt{729 Q^4 \Lambda ^4+4 (3 M \Lambda -3 \Lambda )^3}-27 Q^2 \Lambda ^2\right)^{4/3}}
\end{aligned}$}
\end{equation}
\begin{equation}
\resizebox{1\hsize}{!}{$
\begin{aligned}
\frac{d^2 {S}}{d\Lambda d\Lambda} = -\frac{2^{2/3} \sqrt[3]{\sqrt{729 Q^4 \Lambda ^4+4 (3 M \Lambda -3 \Lambda )^3}-27 Q^2 \Lambda ^2}}{3 \Lambda ^3}+\frac{2^{2/3} \left(\frac{2916 \Lambda ^3 Q^4+12 (3 M-3) (3 M \Lambda -3 \Lambda )^2}{2 \sqrt{729 Q^4 \Lambda ^4+4 (3 M \Lambda -3 \Lambda )^3}}-54 Q^2 \Lambda \right)^2}{27 \Lambda  \left(\sqrt{729 Q^4 \Lambda ^4+4 (3 M \Lambda -3 \Lambda )^3}-27 Q^2 \Lambda ^2\right)^{5/3}}+\\
\frac{4 \sqrt[3]{2} (3 M \Lambda -3 \Lambda ) \left(\frac{2916 \Lambda ^3 Q^4+12 (3 M-3) (3 M \Lambda -3 \Lambda )^2}{2 \sqrt{729 Q^4 \Lambda ^4+4 (3 M \Lambda -3 \Lambda )^3}}-54 Q^2 \Lambda \right)^2}{27 \Lambda  \left(\sqrt{729 Q^4 \Lambda ^4+4 (3 M \Lambda -3 \Lambda )^3}-27 Q^2 \Lambda ^2\right)^{7/3}}+\frac{2^{2/3} \left(\frac{2916 \Lambda ^3 Q^4+12 (3 M-3) (3 M \Lambda -3 \Lambda )^2}{2 \sqrt{729 Q^4 \Lambda ^4+4 (3 M \Lambda -3 \Lambda )^3}}-
54 Q^2 \Lambda \right)}{9 \Lambda ^2 \left(\sqrt{729 Q^4 \Lambda ^4+4 (3 M \Lambda -3 \Lambda )^3}-27 Q^2 \Lambda ^2\right)^{2/3}}+\frac{2 \sqrt[3]{2} (3 M \Lambda -3 \Lambda ) \left(\frac{2916 \Lambda ^3 Q^4+12 (3 M-3) (3 M \Lambda -3 \Lambda )^2}{2 \sqrt{729 Q^4 \Lambda ^4+4 (3 M \Lambda -3 \Lambda )^3}}-
54 Q^2 \Lambda \right)}{9 \Lambda ^2 \left(\sqrt{729 Q^4 \Lambda ^4+4 (3 M \Lambda -3 \Lambda )^3}-27 Q^2 \Lambda ^2\right)^{4/3}}-\\
\frac{2 \sqrt[3]{2} (3 M-3)}{3 \Lambda ^2 \sqrt[3]{\sqrt{729 Q^4 \Lambda ^4+4 (3 M \Lambda -3 \Lambda )^3}-27 Q^2 \Lambda ^2}}+\frac{2 \sqrt[3]{2} (3 M \Lambda -3 \Lambda )}{3 \Lambda ^3 \sqrt[3]{\sqrt{729 Q^4 \Lambda ^4+4 (3 M \Lambda -3 \Lambda )^3}-27 Q^2 \Lambda ^2}}-\\
\frac{-54 Q^2-\frac{\left(2916 \Lambda ^3 Q^4+12 (3 M-3) (3 M \Lambda -3 \Lambda )^2\right)^2}{4 \left(729 Q^4 \Lambda ^4+4 (3 M \Lambda -3 \Lambda )^3\right)^{3/2}}+\frac{8748 \Lambda ^2 Q^4+24 (3 M-3)^2 (3 M \Lambda -3 \Lambda )}{2 \sqrt{729 Q^4 \Lambda ^4+4 (3 M \Lambda -3 \Lambda )^3}}}{9 \sqrt[3]{2} \Lambda  \left(\sqrt{729 Q^4 \Lambda ^4+4 (3 M \Lambda -3 \Lambda )^3}-27 Q^2 \Lambda ^2\right)^{2/3}}-\frac{\sqrt[3]{2} (3 M \Lambda -3 \Lambda ) \left(-54 Q^2-\frac{\left(2916 \Lambda ^3 Q^4+12 (3 M-3) (3 M \Lambda -3 \Lambda )^2\right)^2}{4 \left(729 Q^4 \Lambda ^4+4 (3 M \Lambda -3 \Lambda )^3\right)^{3/2}}+\frac{8748 \Lambda ^2 Q^4+24 (3 M-3)^2 (3 M \Lambda -3 \Lambda )}{2 \sqrt{729 Q^4 \Lambda ^4+4 (3 M \Lambda -3 \Lambda )^3}}\right)}{9 \Lambda  \left(\sqrt{729 Q^4 \Lambda ^4+4 (3 M \Lambda -3 \Lambda )^3}-27 Q^2 \Lambda ^2\right)^{4/3}}-\\
\frac{2 \sqrt[3]{2} (3 M-3) \left(\frac{2916 \Lambda ^3 Q^4+12 (3 M-3) (3 M \Lambda -3 \Lambda )^2}{2 \sqrt{729 Q^4 \Lambda ^4+4 (3 M \Lambda -3 \Lambda )^3}}-54 Q^2 \Lambda \right)}{9 \Lambda  \left(\sqrt{729 Q^4 \Lambda ^4+4 (3 M \Lambda -3 \Lambda )^3}-27 Q^2 \Lambda ^2\right)^{4/3}}
\end{aligned}$}
\end{equation}

The curvature scalar of the thermodynamic geometry can be calculated as:
\begin{equation}
R({S})\rightarrow 1/(\left(4 (3 \Lambda  M-3 \Lambda )^3+729 \Lambda ^4 Q^4\right) \left(\sqrt{4 (3 \Lambda  M-3 \Lambda )^3+729 \Lambda ^4 Q^4}-27 \Lambda ^2 Q^2\right)).
\end{equation}
When $(\left(4 (3 \Lambda  M-3 \Lambda )^3+729 \Lambda ^4 Q^4\right) \left(\sqrt{4 (3 \Lambda  M-3 \Lambda )^3+729 \Lambda ^4 Q^4}-27 \Lambda ^2 Q^2\right))=0$, $R(S)$ diverges, indicating the presence of phase transitions.
\begin{figure}
\centering
\includegraphics[width=0.7\textwidth]{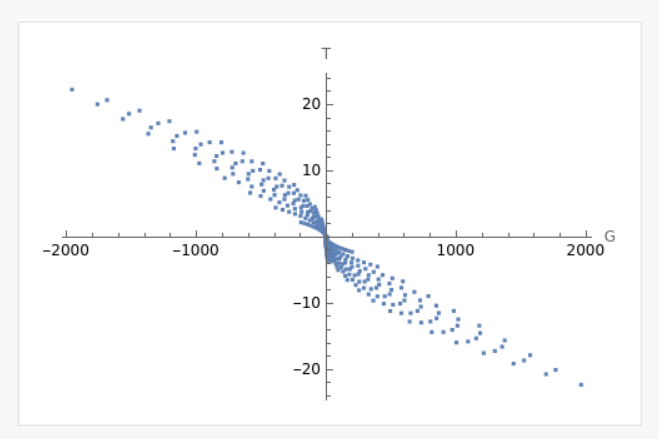}
\caption{At this time, Q=1,$\Lambda$ is transcendental number. $G_4$ is the horizontal axis, and $T_4$ is the vertical axis.}
\label{4.11.png}
\end{figure}
From the plot(section 3.2.4-FIG.17), it can be seen that the $G-T$ curve forms different shapes when $\Lambda$ and $r_{+}$ take different values. In some combinations of $\Lambda$ and $r_{+}$, the shape of the curve is similar to a swallowtail. These situations usually occur when the values of $\Lambda$ and $r_{+}$cause $G_4$ and $T_4$ to vary greatly within a certain range.

\begin{figure}
\centering
\includegraphics[width=0.7\textwidth]{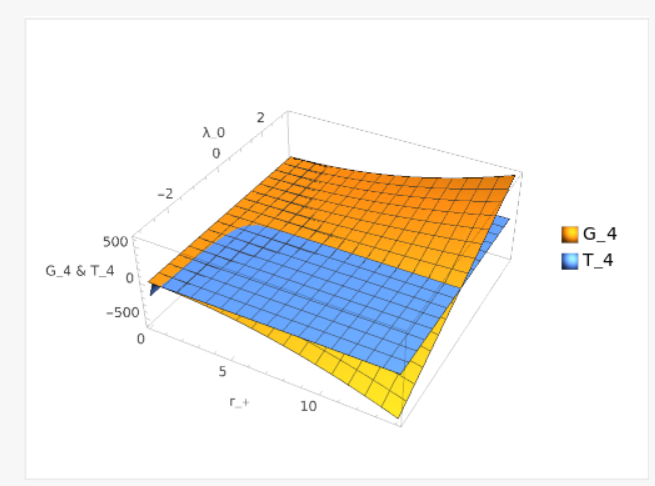}
\caption{At this time, Q=1,$\Lambda_0$ is $\pi$. }
\label{4.111.jpg}
\end{figure}
In the plot (Section 3.2.4-FIG.18), the $G-T$ curve exhibits varying shapes depending on the values of $\Lambda_0$ and $r_{+}$. For certain combinations of $\Lambda_0$ and $r_{+}$, the curve bears a resemblance to a swallowtail. Such configurations typically arise when the values of $\Lambda_0$ and $r_{+}$induce significant variations in $G_4$ and $T_4$ within a specific range.

\subsubsection{When d=3, ${k_1}$=-1}
We have the following equations:

\begin{equation}
g(r)=-\Lambda_0 r^{2}-M +\frac{Q^{2}}{r}-1.
\end{equation}

At this point, the Ricci scalar is obtained by setting $g(r)=0$:

\begin{equation}
\begin{aligned}
& \left\{\left\{r_+ = \frac{2^{1 / 3}(3 \Lambda+3 M \Lambda)}{3 \Lambda\left(-27 Q^2 \Lambda^2+\sqrt{729 Q^4 \Lambda^4+4(3 \Lambda+3 M \Lambda)^3}\right)^{1 / 3}}-\right.\right. \\
& \left.\frac{\left(-27 Q^2 \Lambda^2+\sqrt{729 Q^4 \Lambda^4+4(3 \Lambda+3 M \Lambda)^3}\right)^{1 / 3}}{3 \times 2^{1 / 3} \Lambda}\right\}, \\
&
\end{aligned}
\end{equation}
where $r_+$ is the event horizon radius and the unique Killing horizon radius.

When $g(r)=0$, we get
\begin{equation}
M_5=-\Lambda_0 r^{2} +\frac{Q^{2}}{r}-1.
\end{equation}

The calculation of the Hawking temperature using the conventional method is as follows:
\begin{equation}
T_5=-\frac{\Lambda_0 r_{+}}{2 \pi}-\frac{2Q^{2}}{ r_{+}^{2}}.
\end{equation}

The Gibbs free energy can be derived as:

\begin{equation}
G_5=\Lambda_0 r_{+}^{2}.
\end{equation}

The entropy for this BTZ-f(R) black hole solution is:

\begin{equation}
S=4 \pi r_{+}.
\end{equation}

\begin{figure}
\centering
\includegraphics[width=0.5\textwidth]{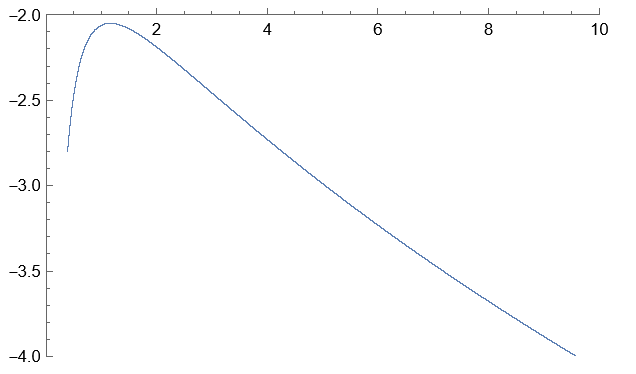}
\caption{At this time, Q=1, $\Lambda$=0.4. $G_5$ is the horizontal axis, and $T_5$ is the vertical axis.}
\label{fig:10}
\end{figure}

\begin{figure}
\centering
\includegraphics[width=0.5\textwidth]{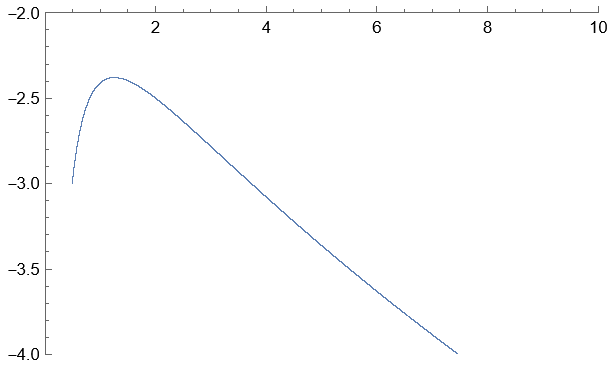}
\caption{At this time, Q=1, $\Lambda_0$=0.5. $G_5$ is the horizontal axis, and $T_5$ is the vertical axis.}
\label{fig:11}
\end{figure}

We can express the metric as:

\begin{equation}
ds^{2}=-\frac{\partial^{2} S({M_5},\Lambda)}{\partial X^{\alpha} \partial X^{\beta}} \Delta X^{\alpha} \Delta X^{\beta},
\end{equation}

where
\begin{equation}
\resizebox{1\hsize}{!}{$
\begin{aligned}
\frac{d^2 {S}}{d{M_5}d {M_5}} ==\frac{18\ 2^{2/3} \Lambda  (3 \Lambda +3 \Lambda  M)^4}{\left(4 (3 \Lambda +3 \Lambda  M)^3+729 \Lambda ^4 Q^4\right)^{3/2} \left(\sqrt{4 (3 \Lambda +3 \Lambda  M)^3+729 \Lambda ^4 Q^4}-27 \Lambda ^2 Q^2\right)^{2/3}}-\\
\frac{6\ 2^{2/3} \Lambda  (3 \Lambda +3 \Lambda  M)}{\sqrt{4 (3 \Lambda +3 \Lambda  M)^3+729 \Lambda ^4 Q^4} \left(\sqrt{4 (3 \Lambda +3 \Lambda  M)^3+729 \Lambda ^4 Q^4}-27 \Lambda ^2 Q^2\right)^{2/3}}+\frac{36 \sqrt[3]{2} \Lambda  (3 \Lambda +3 \Lambda  M)^5}{\left(4 (3 \Lambda +3 \Lambda  M)^3+729 \Lambda ^4 Q^4\right)^{3/2} \left(\sqrt{4 (3 \Lambda +3 \Lambda  M)^3+729 \Lambda ^4 Q^4}-27 \Lambda ^2 Q^2\right)^{4/3}}-\\
\frac{24 \sqrt[3]{2} \Lambda  (3 \Lambda +3 \Lambda  M)^2}{\sqrt{4 (3 \Lambda +3 \Lambda  M)^3+729 \Lambda ^4 Q^4} \left(\sqrt{4 (3 \Lambda +3 \Lambda  M)^3+729 \Lambda ^4 Q^4}-27 \Lambda ^2 Q^2\right)^{4/3}}+\frac{12\ 2^{2/3} \Lambda  (3 \Lambda +3 \Lambda  M)^4}{\left(4 (3 \Lambda +3 \Lambda  M)^3+729 \Lambda ^4 Q^4\right) \left(\sqrt{4 (3 \Lambda +3 \Lambda  M)^3+729 \Lambda ^4 Q^4}-27 \Lambda ^2 Q^2\right)^{5/3}}+\\
\frac{48 \sqrt[3]{2} \Lambda  (3 \Lambda +3 \Lambda  M)^5}{\left(4 (3 \Lambda +3 \Lambda  M)^3+729 \Lambda ^4 Q^4\right) \left(\sqrt{4 (3 \Lambda +3 \Lambda  M)^3+729 \Lambda ^4 Q^4}-27 \Lambda ^2 Q^2\right)^{7/3}}
\end{aligned}$}
\end{equation}
\begin{equation}
\resizebox{1\hsize}{!}{$
\begin{aligned}
\frac{d^2 {S}}{d{M_5}d \Lambda} = \frac{2\ 2^{2/3} (3 M \Lambda +3 \Lambda )^2 \left(\frac{2916 \Lambda ^3 Q^4+12 (3 M+3) (3 M \Lambda +3 \Lambda )^2}{2 \sqrt{729 Q^4 \Lambda ^4+4 (3 M \Lambda +3 \Lambda )^3}}-54 Q^2 \Lambda \right)}{3 \sqrt{729 Q^4 \Lambda ^4+4 (3 M \Lambda +3 \Lambda )^3} \left(\sqrt{729 Q^4 \Lambda ^4+4 (3 M \Lambda +3 \Lambda )^3}-27 Q^2 \Lambda ^2\right)^{5/3}}+\frac{8 \sqrt[3]{2} (3 M \Lambda +3 \Lambda )^3 \left(\frac{2916 \Lambda ^3 Q^4+12 (3 M+3) (3 M \Lambda +3 \Lambda )^2}{2 \sqrt{729 Q^4 \Lambda ^4+4 (3 M \Lambda +3 \Lambda )^3}}-54 Q^2 \Lambda \right)}{3 \sqrt{729 Q^4 \Lambda ^4+4 (3 M \Lambda +3 \Lambda )^3} \left(\sqrt{729 Q^4 \Lambda ^4+4 (3 M \Lambda +3 \Lambda )^3}-27 Q^2 \Lambda ^2\right)^{7/3}}-\\
\frac{\frac{36 (3 M \Lambda +3 \Lambda )^2+72 (3 M+3) \Lambda  (3 M \Lambda +3 \Lambda )}{2 \sqrt{729 Q^4 \Lambda ^4+4 (3 M \Lambda +3 \Lambda )^3}}-\frac{9 \Lambda  (3 M \Lambda +3 \Lambda )^2 \left(2916 \Lambda ^3 Q^4+12 (3 M+3) (3 M \Lambda +3 \Lambda )^2\right)}{\left(729 Q^4 \Lambda ^4+4 (3 M \Lambda +3 \Lambda )^3\right)^{3/2}}}{9 \sqrt[3]{2} \Lambda  \left(\sqrt{729 Q^4 \Lambda ^4+4 (3 M \Lambda +3 \Lambda )^3}-27 Q^2 \Lambda ^2\right)^{2/3}}+\frac{2^{2/3} (3 M \Lambda +3 \Lambda )^2}{\Lambda  \sqrt{729 Q^4 \Lambda ^4+4 (3 M \Lambda +3 \Lambda )^3} \left(\sqrt{729 Q^4 \Lambda ^4+4 (3 M \Lambda +3 \Lambda )^3}-27 Q^2 \Lambda ^2\right)^{2/3}}-\\
\frac{\sqrt[3]{2} \left(\frac{2916 \Lambda ^3 Q^4+12 (3 M+3) (3 M \Lambda +3 \Lambda )^2}{2 \sqrt{729 Q^4 \Lambda ^4+4 (3 M \Lambda +3 \Lambda )^3}}-54 Q^2 \Lambda \right)}{3 \left(\sqrt{729 Q^4 \Lambda ^4+4 (3 M \Lambda +3 \Lambda )^3}-27 Q^2 \Lambda ^2\right)^{4/3}}-\frac{\sqrt[3]{2} (3 M \Lambda +3 \Lambda ) \left(\frac{36 (3 M \Lambda +3 \Lambda )^2+72 (3 M+3) \Lambda  (3 M \Lambda +3 \Lambda )}{2 \sqrt{729 Q^4 \Lambda ^4+4 (3 M \Lambda +3 \Lambda )^3}}-\frac{9 \Lambda  (3 M \Lambda +3 \Lambda )^2 \left(2916 \Lambda ^3 Q^4+12 (3 M+3) (3 M \Lambda +3 \Lambda )^2\right)}{\left(729 Q^4 \Lambda ^4+4 (3 M \Lambda +3 \Lambda )^3\right)^{3/2}}\right)}{9 \Lambda  \left(\sqrt{729 Q^4 \Lambda ^4+4 (3 M \Lambda +3 \Lambda )^3}-27 Q^2 \Lambda ^2\right)^{4/3}}-\\
\frac{2 \sqrt[3]{2} (3 M+3) (3 M \Lambda +3 \Lambda )^2}{\sqrt{729 Q^4 \Lambda ^4+4 (3 M \Lambda +3 \Lambda )^3} \left(\sqrt{729 Q^4 \Lambda ^4+4 (3 M \Lambda +3 \Lambda )^3}-27 Q^2 \Lambda ^2\right)^{4/3}}+\frac{2 \sqrt[3]{2} (3 M \Lambda +3 \Lambda )^3}{\Lambda  \sqrt{729 Q^4 \Lambda ^4+4 (3 M \Lambda +3 \Lambda )^3} \left(\sqrt{729 Q^4 \Lambda ^4+4 (3 M \Lambda +3 \Lambda )^3}-27 Q^2 \Lambda ^2\right)^{4/3}}
\end{aligned}$}
\end{equation}
\begin{equation}
\resizebox{1\hsize}{!}{$
\begin{aligned}
\frac{d^2 {S}}{d \Lambda d \Lambda} =-\frac{2^{2/3} \sqrt[3]{\sqrt{729 Q^4 \Lambda ^4+4 (3 M \Lambda +3 \Lambda )^3}-27 Q^2 \Lambda ^2}}{3 \Lambda ^3}+\frac{2^{2/3} \left(\frac{2916 \Lambda ^3 Q^4+12 (3 M+3) (3 M \Lambda +3 \Lambda )^2}{2 \sqrt{729 Q^4 \Lambda ^4+4 (3 M \Lambda +3 \Lambda )^3}}-54 Q^2 \Lambda \right)^2}{27 \Lambda  \left(\sqrt{729 Q^4 \Lambda ^4+4 (3 M \Lambda +3 \Lambda )^3}-27 Q^2 \Lambda ^2\right)^{5/3}}+\\
\frac{4 \sqrt[3]{2} (3 M \Lambda +3 \Lambda ) \left(\frac{2916 \Lambda ^3 Q^4+12 (3 M+3) (3 M \Lambda +3 \Lambda )^2}{2 \sqrt{729 Q^4 \Lambda ^4+4 (3 M \Lambda +3 \Lambda )^3}}-54 Q^2 \Lambda \right)^2}{27 \Lambda  \left(\sqrt{729 Q^4 \Lambda ^4+4 (3 M \Lambda +3 \Lambda )^3}-27 Q^2 \Lambda ^2\right)^{7/3}}+\frac{2^{2/3} \left(\frac{2916 \Lambda ^3 Q^4+12 (3 M+3) (3 M \Lambda +3 \Lambda )^2}{2 \sqrt{729 Q^4 \Lambda ^4+4 (3 M \Lambda +3 \Lambda )^3}}-54 Q^2 \Lambda \right)}{9 \Lambda ^2 \left(\sqrt{729 Q^4 \Lambda ^4+4 (3 M \Lambda +3 \Lambda )^3}-27 Q^2 \Lambda ^2\right)^{2/3}}+\frac{2 \sqrt[3]{2} (3 M \Lambda +3 \Lambda ) \left(\frac{2916 \Lambda ^3 Q^4+12 (3 M+3) (3 M \Lambda +3 \Lambda )^2}{2 \sqrt{729 Q^4 \Lambda ^4+4 (3 M \Lambda +3 \Lambda )^3}}-54 Q^2 \Lambda \right)}{9 \Lambda ^2 \left(\sqrt{729 Q^4 \Lambda ^4+4 (3 M \Lambda +3 \Lambda )^3}-27 Q^2 \Lambda ^2\right)^{4/3}}-\\
\frac{2 \sqrt[3]{2} (3 M+3)}{3 \Lambda ^2 \sqrt[3]{\sqrt{729 Q^4 \Lambda ^4+4 (3 M \Lambda +3 \Lambda )^3}-27 Q^2 \Lambda ^2}}+\frac{2 \sqrt[3]{2} (3 M \Lambda +3 \Lambda )}{3 \Lambda ^3 \sqrt[3]{\sqrt{729 Q^4 \Lambda ^4+4 (3 M \Lambda +3 \Lambda )^3}-27 Q^2 \Lambda ^2}}-\frac{-54 Q^2-\frac{\left(2916 \Lambda ^3 Q^4+12 (3 M+3) (3 M \Lambda +3 \Lambda )^2\right)^2}{4 \left(729 Q^4 \Lambda ^4+4 (3 M \Lambda +3 \Lambda )^3\right)^{3/2}}+\frac{8748 \Lambda ^2 Q^4+24 (3 M+3)^2 (3 M \Lambda +3 \Lambda )}{2 \sqrt{729 Q^4 \Lambda ^4+4 (3 M \Lambda +3 \Lambda )^3}}}{9 \sqrt[3]{2} \Lambda  \left(\sqrt{729 Q^4 \Lambda ^4+4 (3 M \Lambda +3 \Lambda )^3}-27 Q^2 \Lambda ^2\right)^{2/3}}-\\
\frac{\sqrt[3]{2} (3 M \Lambda +3 \Lambda ) \left(-54 Q^2-\frac{\left(2916 \Lambda ^3 Q^4+12 (3 M+3) (3 M \Lambda +3 \Lambda )^2\right)^2}{4 \left(729 Q^4 \Lambda ^4+4 (3 M \Lambda +3 \Lambda )^3\right)^{3/2}}+\frac{8748 \Lambda ^2 Q^4+24 (3 M+3)^2 (3 M \Lambda +3 \Lambda )}{2 \sqrt{729 Q^4 \Lambda ^4+4 (3 M \Lambda +3 \Lambda )^3}}\right)}{9 \Lambda  \left(\sqrt{729 Q^4 \Lambda ^4+4 (3 M \Lambda +3 \Lambda )^3}-27 Q^2 \Lambda ^2\right)^{4/3}}-\frac{2 \sqrt[3]{2} (3 M+3) \left(\frac{2916 \Lambda ^3 Q^4+12 (3 M+3) (3 M \Lambda +3 \Lambda )^2}{2 \sqrt{729 Q^4 \Lambda ^4+4 (3 M \Lambda +3 \Lambda )^3}}-54 Q^2 \Lambda \right)}{9 \Lambda  \left(\sqrt{729 Q^4 \Lambda ^4+4 (3 M \Lambda +3 \Lambda )^3}-27 Q^2 \Lambda ^2\right)^{4/3}}
\end{aligned}$}
\end{equation}

The curvature scalar of the thermodynamic geometry can be calculated as:
\begin{equation}
R({S})\rightarrow 1/(\left(4 (3 \Lambda  M+3 \Lambda )^3+729 \Lambda ^4 Q^4\right) \left(\sqrt{4 (3 \Lambda  M+3 \Lambda )^3+729 \Lambda ^4 Q^4}-27 \Lambda ^2 Q^2\right)).
\end{equation}
When $(\left(4 (3 \Lambda  M+3 \Lambda )^3+729 \Lambda ^4 Q^4\right) \left(\sqrt{4 (3 \Lambda  M+3 \Lambda )^3+729 \Lambda ^4 Q^4}-27 \Lambda ^2 Q^2\right))=0$, $R(S)$ diverges, indicating the presence of phase transitions.

\begin{figure}
\centering
\includegraphics[width=0.7\textwidth]{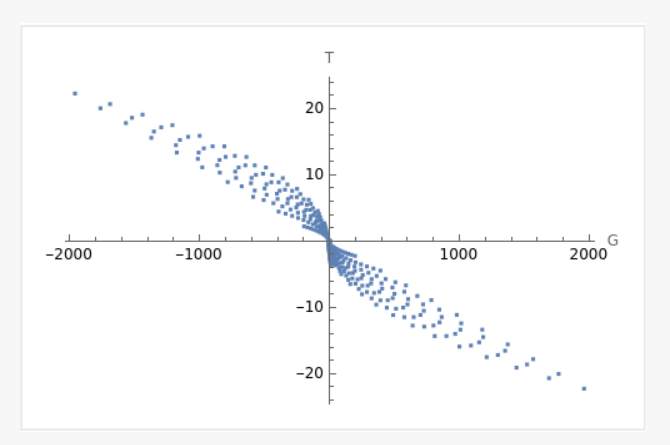}
\caption{At this time, Q=1,$\Lambda$ is transcendental number. $G_5$ is the horizontal axis, and $T_5$ is the vertical axis.}
\label{5.11.png}
\end{figure}From FIG.21, it can be seen that the $G-T$ graph will show different shapes, including a swallowtail shape, when $\Lambda$ and $r_{+}$take different values. This implies that phase transitions will occur in the system under certain values of $\Lambda$ and $r_{+}$, which is the physical meaning of the swallowtail shape.

\begin{figure}
\centering
\includegraphics[width=0.7\textwidth]{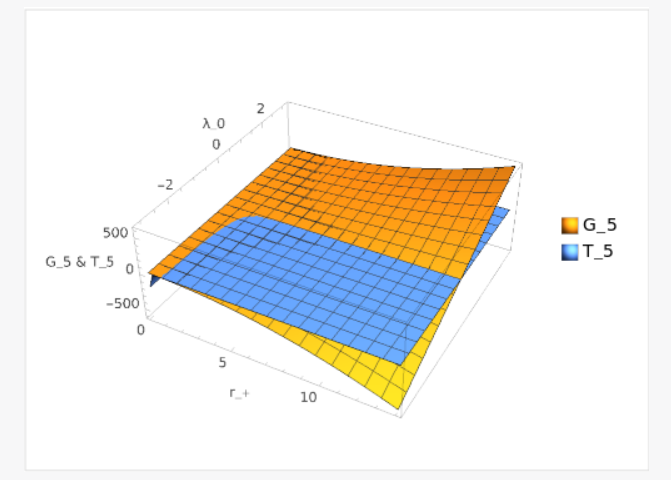}
\caption{At this time, Q=1,$\Lambda_0$ is $\pi$.}
\label{5.111.jpg}
\end{figure}In the plot (Section 3.2.5-(FIG.22)), the $G-T$ graph manifests various shapes, notably the distinctive swallowtail, depending on the values of $\Lambda_0$ and $r_{+}$. This suggests that phase transitions are likely to occur in the system for specific combinations of $\Lambda_0$ and $r_{+}$. The emergence of the swallowtail shape serves as a physical representation of these transitions.

\subsubsection{When d=3, ${k_1}$=0}
We have the following equations:

\begin{equation}
g(r)=-\Lambda_0 r^{2}-M +\frac{Q^{2}}{r}.
\end{equation}

At this point, the Ricci scalar is obtained by setting $g(r)=0$:

\begin{equation}
\begin{aligned}
\left\{r_+= \frac{\sqrt[3]{\frac{2}{3}} M}{\sqrt[3]{\sqrt{3} \sqrt{4 \Lambda ^3 M^3+27 \Lambda ^4 Q^4}-9 \Lambda ^2 Q^2}}-\frac{\sqrt[3]{\sqrt{3} \sqrt{4 \Lambda ^3 M^3+27 \Lambda ^4 Q^4}-9 \Lambda ^2 Q^2}}{\sqrt[3]{2}* 3^{2/3} \Lambda }\right\}
\end{aligned}
\end{equation}

where $r_+$ is the event horizon radius and the unique Killing horizon radius.

When $g(r)=0$, we get

\begin{equation}
M=M_6=-\Lambda_0 r^{2} +\frac{Q^{2}}{r}.
\end{equation}

The calculation of the Hawking temperature using the conventional method is as follows:

\begin{equation}
T_6=-\frac{\Lambda_0 r_{+}}{2 \pi}-\frac{2Q^{2}}{ r_{+}^{2}}.
\end{equation}

The Gibbs free energy can be derived as:

\begin{equation}
G_6=\Lambda_0 r_{+}^{2}.
\end{equation}

The entropy for this BTZ-f(R) black hole solution is:

\begin{equation}
S=4 \pi r_{+}.
\end{equation}

\begin{figure}
\centering
\includegraphics[width=0.5\textwidth]{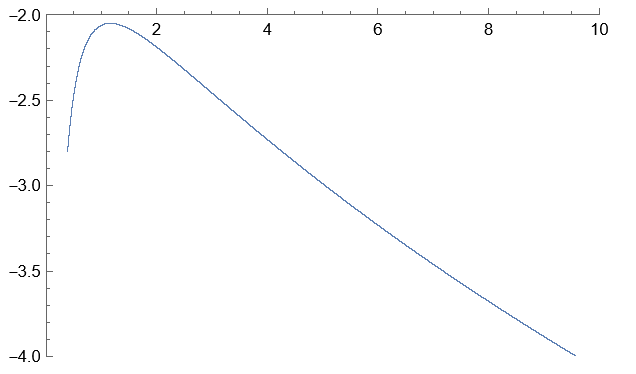}
\caption{At this time, Q=1, $\Lambda_0$=0.4. $G_6$ is the horizontal axis, and $T_6$ is the vertical axis.}
\label{fig:10}
\end{figure}

\begin{figure}
\centering
\includegraphics[width=0.5\textwidth]{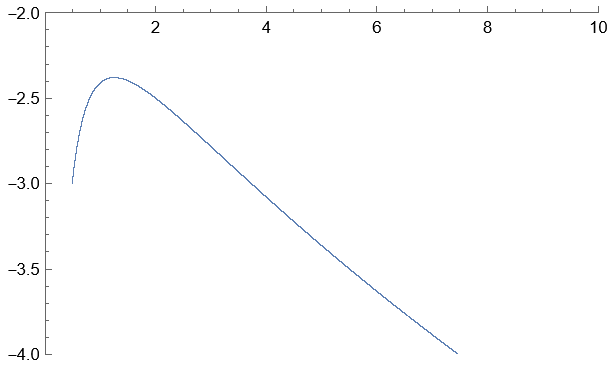}
\caption{At this time, Q=1, $\Lambda_0$=0.5. $G_6$ is the horizontal axis, and $T_6$ is the vertical axis.}
\label{fig:11}
\end{figure}

We can express the metric as:

\begin{equation}
ds^{2}=-\frac{\partial^{2} S({M_6},\Lambda)}{\partial X^{\alpha} \partial X^{\beta}} \Delta X^{\alpha} \Delta X^{\beta},
\end{equation}

where
\begin{equation}
\resizebox{\textwidth}{!}{$
\begin{aligned}
\frac{d^2 {S}}{d{M_6}d {M_6}} =-\frac{2\ 2^{2/3} \Lambda ^2 M}{\sqrt[6]{3} \sqrt{4 \Lambda ^3 M^3+27 \Lambda ^4 Q^4} \left(\sqrt{3} \sqrt{4 \Lambda ^3 M^3+27 \Lambda ^4 Q^4}-9 \Lambda ^2 Q^2\right)^{2/3}}+\frac{12 \sqrt[3]{2} \sqrt[6]{3} \Lambda ^6 M^5}{\left(4 \Lambda ^3 M^3+27 \Lambda ^4 Q^4\right)^{3/2} \left(\sqrt{3} \sqrt{4 \Lambda ^3 M^3+27 \Lambda ^4 Q^4}-9 \Lambda ^2 Q^2\right)^{4/3}}+\\
\frac{16 \sqrt[3]{2} 3^{2/3} \Lambda ^6 M^5}{\left(4 \Lambda ^3 M^3+27 \Lambda ^4 Q^4\right) \left(\sqrt{3} \sqrt{4 \Lambda ^3 M^3+27 \Lambda ^4 Q^4}-9 \Lambda ^2 Q^2\right)^{7/3}}+\frac{2\ 2^{2/3} 3^{5/6} \Lambda ^5 M^4}{\left(4 \Lambda ^3 M^3+27 \Lambda ^4 Q^4\right)^{3/2} \left(\sqrt{3} \sqrt{4 \Lambda ^3 M^3+27 \Lambda ^4 Q^4}-9 \Lambda ^2 Q^2\right)^{2/3}}+\\
\frac{4\ 2^{2/3} \sqrt[3]{3} \Lambda ^5 M^4}{\left(4 \Lambda ^3 M^3+27 \Lambda ^4 Q^4\right) \left(\sqrt{3} \sqrt{4 \Lambda ^3 M^3+27 \Lambda ^4 Q^4}-9 \Lambda ^2 Q^2\right)^{5/3}}-\frac{8 \sqrt[3]{2} \sqrt[6]{3} \Lambda ^3 M^2}{\sqrt{4 \Lambda ^3 M^3+27 \Lambda ^4 Q^4} \left(\sqrt{3} \sqrt{4 \Lambda ^3 M^3+27 \Lambda ^4 Q^4}-9 \Lambda ^2 Q^2\right)^{4/3}}
\end{aligned}$}
\end{equation}
\begin{equation}
\resizebox{\textwidth}{!}{$
\begin{aligned}
\frac{d^2 {S}}{d{M_6}d \Lambda} = -\frac{\sqrt[3]{\frac{2}{3}} \left(\frac{\sqrt{3} \left(12 \Lambda ^2 M^3+108 \Lambda ^3 Q^4\right)}{2 \sqrt{4 \Lambda ^3 M^3+27 \Lambda ^4 Q^4}}-18 \Lambda  Q^2\right)}{3 \left(\sqrt{3} \sqrt{4 \Lambda ^3 M^3+27 \Lambda ^4 Q^4}-9 \Lambda ^2 Q^2\right)^{4/3}}+\frac{8 \sqrt[3]{2} \Lambda ^3 M^3 \left(\frac{\sqrt{3} \left(12 \Lambda ^2 M^3+108 \Lambda ^3 Q^4\right)}{2 \sqrt{4 \Lambda ^3 M^3+27 \Lambda ^4 Q^4}}-18 \Lambda  Q^2\right)}{3^{5/6} \sqrt{4 \Lambda ^3 M^3+27 \Lambda ^4 Q^4} \left(\sqrt{3} \sqrt{4 \Lambda ^3 M^3+27 \Lambda ^4 Q^4}-9 \Lambda ^2 Q^2\right)^{7/3}}-\\
\frac{\sqrt[3]{\frac{2}{3}} M \left(\frac{18 \sqrt{3} \Lambda ^2 M^2}{\sqrt{4 \Lambda ^3 M^3+27 \Lambda ^4 Q^4}}-\frac{3 \sqrt{3} \Lambda ^3 M^2 \left(12 \Lambda ^2 M^3+108 \Lambda ^3 Q^4\right)}{\left(4 \Lambda ^3 M^3+27 \Lambda ^4 Q^4\right)^{3/2}}\right)}{3 \left(\sqrt{3} \sqrt{4 \Lambda ^3 M^3+27 \Lambda ^4 Q^4}-9 \Lambda ^2 Q^2\right)^{4/3}}+\\
\frac{2\ 2^{2/3} \Lambda ^2 M^2 \left(\frac{\sqrt{3} \left(12 \Lambda ^2 M^3+108 \Lambda ^3 Q^4\right)}{2 \sqrt{4 \Lambda ^3 M^3+27 \Lambda ^4 Q^4}}-18 \Lambda  Q^2\right)}{3 \sqrt[6]{3} \sqrt{4 \Lambda ^3 M^3+27 \Lambda ^4 Q^4} \left(\sqrt{3} \sqrt{4 \Lambda ^3 M^3+27 \Lambda ^4 Q^4}-9 \Lambda ^2 Q^2\right)^{5/3}}-\frac{\frac{18 \sqrt{3} \Lambda ^2 M^2}{\sqrt{4 \Lambda ^3 M^3+27 \Lambda ^4 Q^4}}-\frac{3 \sqrt{3} \Lambda ^3 M^2 \left(12 \Lambda ^2 M^3+108 \Lambda ^3 Q^4\right)}{\left(4 \Lambda ^3 M^3+27 \Lambda ^4 Q^4\right)^{3/2}}}{3 \sqrt[3]{2} 3^{2/3} \Lambda  \left(\sqrt{3} \sqrt{4 \Lambda ^3 M^3+27 \Lambda ^4 Q^4}-9 \Lambda ^2 Q^2\right)^{2/3}}+\frac{2^{2/3} \Lambda  M^2}{\sqrt[6]{3} \sqrt{4 \Lambda ^3 M^3+27 \Lambda ^4 Q^4} \left(\sqrt{3} \sqrt{4 \Lambda ^3 M^3+27 \Lambda ^4 Q^4}-9 \Lambda ^2 Q^2\right)^{2/3}}
\end{aligned}$}
\end{equation}
\begin{equation}
\resizebox{\textwidth}{!}{$
\begin{aligned}
\frac{d^2 {S}}{d\Lambda d\Lambda} = \frac{4 \sqrt[3]{\frac{2}{3}} M \left(\frac{\sqrt{3} \left(12 \Lambda ^2 M^3+108 \Lambda ^3 Q^4\right)}{2 \sqrt{4 \Lambda ^3 M^3+27 \Lambda ^4 Q^4}}-18 \Lambda  Q^2\right)^2}{9 \left(\sqrt{3} \sqrt{4 \Lambda ^3 M^3+27 \Lambda ^4 Q^4}-9 \Lambda ^2 Q^2\right)^{7/3}}-\frac{\sqrt[3]{\frac{2}{3}} M \left(-\frac{\sqrt{3} \left(12 \Lambda ^2 M^3+108 \Lambda ^3 Q^4\right)^2}{4 \left(4 \Lambda ^3 M^3+27 \Lambda ^4 Q^4\right)^{3/2}}+\frac{\sqrt{3} \left(24 \Lambda  M^3+324 \Lambda ^2 Q^4\right)}{2 \sqrt{4 \Lambda ^3 M^3+27 \Lambda ^4 Q^4}}-18 Q^2\right)}{3 \left(\sqrt{3} \sqrt{4 \Lambda ^3 M^3+27 \Lambda ^4 Q^4}-9 \Lambda ^2 Q^2\right)^{4/3}}+\frac{\left(\frac{2}{3}\right)^{2/3} \left(\frac{\sqrt{3} \left(12 \Lambda ^2 M^3+108 \Lambda ^3 Q^4\right)}{2 \sqrt{4 \Lambda ^3 M^3+27 \Lambda ^4 Q^4}}-18 \Lambda  Q^2\right)^2}{9 \Lambda  \left(\sqrt{3} \sqrt{4 \Lambda ^3 M^3+27 \Lambda ^4 Q^4}-9 \Lambda ^2 Q^2\right)^{5/3}}+\\
\frac{\left(\frac{2}{3}\right)^{2/3} \left(\frac{\sqrt{3} \left(12 \Lambda ^2 M^3+108 \Lambda ^3 Q^4\right)}{2 \sqrt{4 \Lambda ^3 M^3+27 \Lambda ^4 Q^4}}-18 \Lambda  Q^2\right)}{3 \Lambda ^2 \left(\sqrt{3} \sqrt{4 \Lambda ^3 M^3+27 \Lambda ^4 Q^4}-9 \Lambda ^2 Q^2\right)^{2/3}}-\frac{\left(\frac{2}{3}\right)^{2/3} \sqrt[3]{\sqrt{3} \sqrt{4 \Lambda ^3 M^3+27 \Lambda ^4 Q^4}-9 \Lambda ^2 Q^2}}{\Lambda ^3}-\frac{-\frac{\sqrt{3} \left(12 \Lambda ^2 M^3+108 \Lambda ^3 Q^4\right)^2}{4 \left(4 \Lambda ^3 M^3+27 \Lambda ^4 Q^4\right)^{3/2}}+\frac{\sqrt{3} \left(24 \Lambda  M^3+324 \Lambda ^2 Q^4\right)}{2 \sqrt{4 \Lambda ^3 M^3+27 \Lambda ^4 Q^4}}-18 Q^2}{3 \sqrt[3]{2} 3^{2/3} \Lambda  \left(\sqrt{3} \sqrt{4 \Lambda ^3 M^3+27 \Lambda ^4 Q^4}-9 \Lambda ^2 Q^2\right)^{2/3}}
\end{aligned}$}
\end{equation}

The curvature scalar of the thermodynamic geometry can be calculated as:
\begin{equation}
R({S})\rightarrow 1/(\left(4 (3 \Lambda  M-3 \Lambda )^3+729 \Lambda ^4 Q^4\right) \left(\sqrt{4 (3 \Lambda  M-3 \Lambda )^3+729 \Lambda ^4 Q^4}-27 \Lambda ^2 Q^2\right)).
\end{equation}
When $(\left(4 (3 \Lambda  M-3 \Lambda )^3+729 \Lambda ^4 Q^4\right) \left(\sqrt{4 (3 \Lambda  M-3 \Lambda )^3+729 \Lambda ^4 Q^4}-27 \Lambda ^2 Q^2\right))=0$, $R(S)$ diverges, indicating the presence of phase transitions.

The partial derivatives of $S$ with respect to $M_6$ and $\Lambda$ are given by the above equations.
\begin{figure}
\centering
\includegraphics[width=0.7\textwidth]{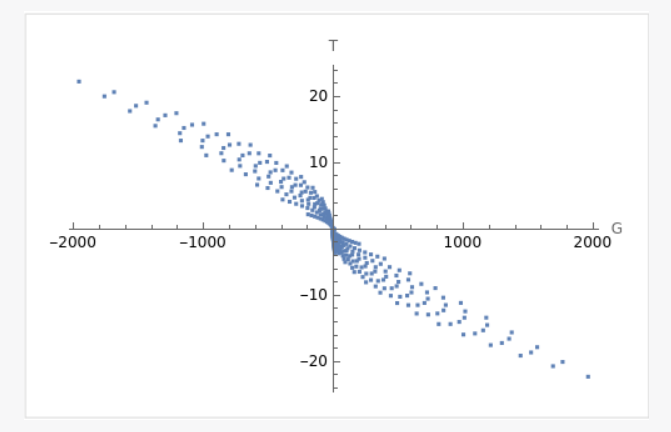}
\caption{At this time, Q=1. $G_6$ is the horizontal axis, and $T_6$ is the vertical axis.This is the plot of function $G$ versus $T_6$, where $G_6$ is the Gibbs free energy, $T_6$ is the temperature, $\Lambda_0$ is a transcendental number about $\pi$,and the range of $r_{+}$is from 1 to 14 .}
\label{6.11.jpg}
\end{figure}From the plot(Section 3.2.6-FIG.25), it can be seen that the $G-T$ graph will show different shapes, including a swallowtail shape, when $\Lambda$ and $r_{+}$take different values. This implies that phase transitions will occur in the system under certain values of $\Lambda$ and $r_{+}$, which is the physical meaning of the swallowtail shape.

\begin{figure}
\centering
\includegraphics[width=0.7\textwidth]{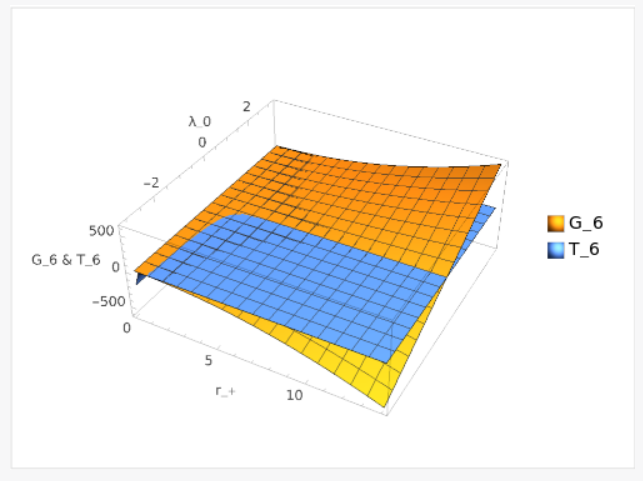}
\caption{The equations for $G_6$ and $T_6$ are the same as the ones provided for $G_4$ and $T_4$ as well as $G_5$ and $T_5$ in the previous queries. Given that $Q=1$ and $\Lambda_0=\pi$, the plot for $G_6$ and $T_6$ will be identical to the ones generated for the previous sets of equations.}
\label{6.111.jpg}
\end{figure}In the plot (Section 3.2.6-FIG.26), the $G-T$ graph exhibits various configurations, with the swallowtail shape being particularly prominent under certain combinations of $\Lambda$ and $r_{+}$. This suggests that the system undergoes phase transitions at specific values of $\Lambda$ and $r_{+}$. The presence of the swallowtail shape provides a physical interpretation of these transitions.

\subsection{Black hole in the form of f(R) theory: $$f(R)=-2 \eta M \ln (6 \Lambda_0+R)+R_{0}$$}
In the case where $\Phi(r)=0$, the charged $(2+1)$-dimensional solution under pure $f(R)$-gravity can be expressed using the metric given in follow,\cite{30,31,32,33,34}, where
\begin{equation}
g(r)=-\Lambda_0 r^{2}-M r-\frac{2 Q^{2}}{3 \eta r}.
\end{equation}

The two-dimensional line element is \cite{10,17}
\begin{equation}
d s^{2}=g(r) d \tau^{2}+\frac{d r^{2}}{g(r)}.
\end{equation}
To ensure consistency, we have made the assumption that both $\eta$ and $\xi$ do not take negative values.

At this point, the Ricci scalar is obtained by setting $g(r)=0$:
\begin{equation}
\begin{aligned}
&r_+=\frac{1}{3}\left(-\frac{M}{\Lambda_0}-\frac{M^{2} \eta}{\Lambda_0\left(M^{3} \eta^{3}+9 Q^{2} \eta^{2} \Lambda_0^{2}+3 \sqrt{2 M^{3} Q^{2} \eta^{5} \Lambda_0^{2}+9 Q^{4} \eta^{4} \Lambda_0^{4}}\right)^{1 / 3}}-\right.\\\
&\left.\frac{\left(M^{3} \eta^{3}+9 Q^{2} \eta^{2} \Lambda_0^{2}+3 \sqrt{2 M^{3} Q^{2} \eta^{5} \Lambda_0^{2}+9 Q^{4} \eta^{4} \Lambda_0^{4}}\right) ^{1 / 3}}{\eta \Lambda_0}\right),
\end{aligned}
\end{equation}where ${r_+}$ is the event horizon radius and the unique Killing horizon radius.

The calculation of the Hawking temperature using the conventional method is as follows:
\begin{equation}
T_7=-\frac{\Lambda_0 r_{+}}{2 \pi}-\frac{M}{4 \pi}+\frac{Q^{2}}{6 \pi \eta r_{+}^{2}}.
\end{equation}

Gibbs free energy can be derived as
\begin{equation}
G_7=\Lambda_0 r_{+}^{2}
\end{equation}
 The entropy for this BTZ-f(R) Black hole solution is
\begin{equation}
S=4 \pi r_{+}.
\end{equation}

\begin{equation}
{G_7}=H-{T_7}S={M_7}-{T_7} S=\frac{3 Q^2}{4 r_{+}}-\frac{2 P \pi r_{+}^3}{3}.
\end{equation}
\begin{figure}
  \centering
  \includegraphics[width=0.5\textwidth]{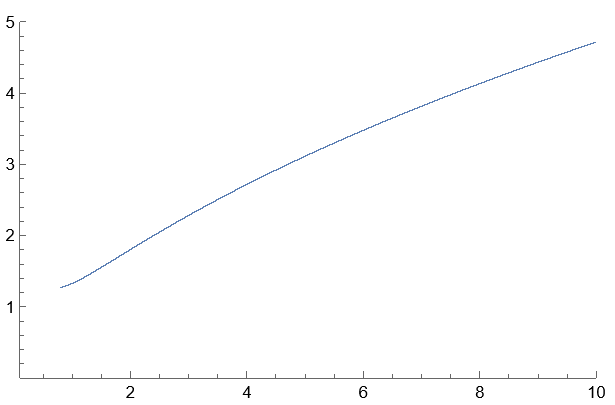}
  \caption{At this time, Q=1, P=0.8,M=1,$\eta=1$,$G_7$ is the horizontal axis, $T_7$ is the vertical axis}
  \label{8.png}
\end{figure}
\begin{figure}
  \centering
  \includegraphics[width=0.5\textwidth]{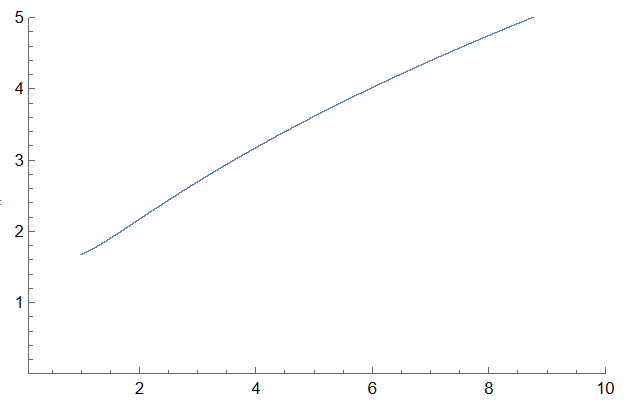}
  \caption{At this time, Q=1, P=1,M=1,$\eta=1$,$G_7$ is the horizontal axis, $T_7$ is the vertical axis}
  \label{9.png}
\end{figure}

We can express the metric as:
\begin{equation}
d s^{2}=-\frac{\partial^{2}  S({M_7},\Lambda_0)}{\partial X^{\alpha} \partial X^{\beta}} \Delta X^{\alpha} \Delta X^{\beta},
\end{equation}
where
\begin{equation}
g_{ij}\rightarrow 1/\left(M^3 \eta^3+9 Q^2 \eta^2 \Lambda_0^2+3 \sqrt{2 M^3 Q^2 \eta^5 \Lambda_0^2+9 Q^4 \eta^4 \Lambda_0^4}\right)
\end{equation}
\begin{equation}
R(S)\rightarrow 1/\left(M^3 \eta^3+9 Q^2 \eta^2 \Lambda_0^2+3 \sqrt{2 M^3 Q^2 \eta^5 \Lambda_0^2+9 Q^4 \eta^4 \Lambda_0^4}\right)
\end{equation}The mentioned above do not involve any phase transition.

\subsection{Black hole in the form of f(R) theory:$f(R)=2 a \sqrt{R-\alpha}$}

The f(R) theory describes a black hole that takes the form of $f(R)=2 a \sqrt{R-\alpha}$, as stated in \cite{18,19,20,21,22,23,24,25,26,27,28,29,30}. In this model, $\alpha$ is a parameter associated with an effective cosmological constant and is expressed in units of [distance]$^{-1}$, with the condition that $\alpha$ is greater than 0. The static spherically symmetric black hole in this model is defined by the following metric:
\begin{equation}
d s^{2}=-g(r) d t^{2}+\frac{d r^{2}}{g(r)}+r^{2} d \Omega^{2},
\end{equation}
where
\begin{equation}
g(r)=\frac{1}{2}\left(1-\frac{\alpha r^{2}}{6}+\frac{2 Q}{r^{2}}\right).
\end{equation}
$Q$ is an integration constant. The black hole's event horizon is located at different radii depending on the values of $\alpha$ and $Q$: $(a)$ when $\alpha>0$ and $Q>0$, the event horizon is at $r_{+}=\sqrt{3 \alpha+\alpha \sqrt{9+12 \alpha Q}}/\alpha$; (b) when $\alpha>0$, $Q<0$ and $\alpha Q>-3/4$, the event horizon is at $r_{+}=\sqrt{3\alpha-\alpha \sqrt{9+12 \alpha Q}}/\alpha$; (c) when $\alpha<0$ and $Q<0$, the event horizon is at $r_{+}=\sqrt{3\alpha-\alpha \sqrt{9+12 \alpha Q}} / \alpha$; (d) when $\alpha>0$ and $Q=0$, the event horizon is at $r_{+}=\sqrt{6/\alpha}$.

In cases where $\alpha>0$ and $Q>0$, or $\alpha<0$ and $Q<0$, or $\alpha>0$ and $Q=0$, there is only one positive root of $g(r)=0$ that has physical significance. 

S can be calculated using the formula:
\begin{equation}
S=\int_0^{r_{+}} \frac{1}{T}\left(\frac{\partial M}{\partial r_{+}}\right) d r_{+}=\pi r_{+}^2.
\end{equation}

When the cosmological constant $\alpha$ is negative
\begin{equation}
P=-\frac{\alpha}{4 \pi} ,\quad V=\left(\frac{\partial M}{\partial P}\right)_{S, q}.
\end{equation}

With the introduction of the newly defined thermodynamic quantities, the first law of black hole thermodynamics in the extended phase space is given by:

\begin{equation}
dM = TdS + \Phi dq + VdP,
\end{equation}
where $M$ is the mass, $T$ is the temperature, $S$ is the entropy, $\Phi$ is the electrostatic potential, $q$ is the electric charge, $V$ is the thermodynamic volume, and $P$ is the pressure.

And 
\begin{equation}
M=2 T S+\Phi q-2 V P.
\end{equation}

Moreover, within the extended phase space, the mass is interpreted as enthalpy. Therefore, the Gibbs free energy can be derived as:
\begin{equation}
{G_8} = H - {T_8}S = M - {T_8}S = \frac{3Q}{2r_+} + \frac{r_+}{4} - \frac{2P\pi r_+^3}{3},
\end{equation}
where $H$ is the enthalpy, and $r_+$ is the black hole event horizon.
\begin{figure}
  \centering
  \includegraphics[width=0.5\textwidth]{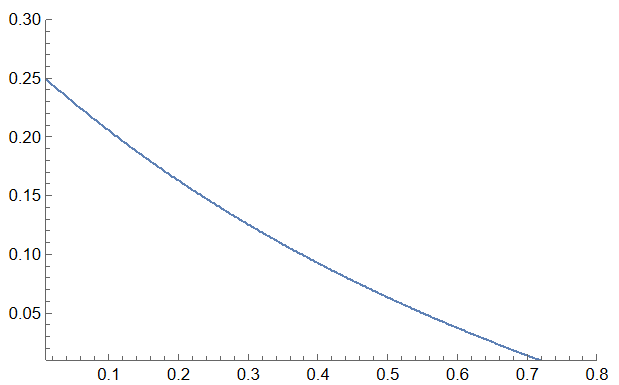}
  \caption{At this time, Q=-1, $\alpha$=0.6, $G_8$ is the horizontal axis, $T_8$ is the vertical axis}
  \label{3.png}
\end{figure}
\begin{figure}
  \centering
  \includegraphics[width=0.5\textwidth]{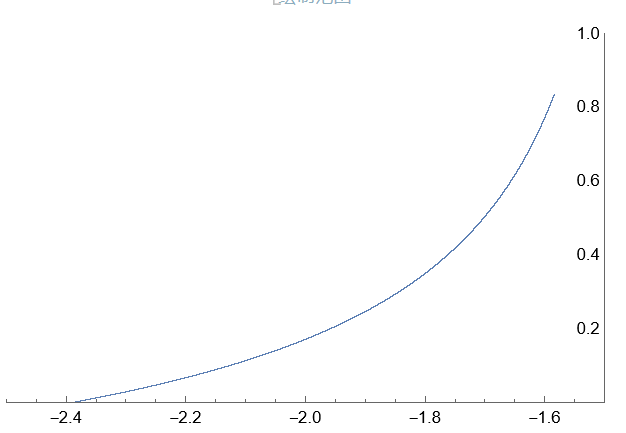}
  \caption{At this time, Q=-1, $\alpha$=1, $G_8$ is the horizontal axis, $T_8$ is the vertical axis}
  \label{3.1.png}
\end{figure}

When ${S_1}=r_{+}^2=(3 \alpha+\alpha \sqrt{9+12 \alpha Q})/(\alpha)^2$,
\begin{equation}
\frac{d^2{S_1}}{d \alpha^2}=\frac{6\left(2 \sqrt{3} \alpha^2 Q^2+3(\sqrt{3}+\sqrt{3+4 \alpha Q})+a Q(6 \sqrt{3}+4 \sqrt{3+4\alpha Q})\right)}{\alpha^3(3+4 \alpha Q)^{3 / 2}},
\end{equation}
\begin{equation}
\frac{d^2 {S_1}}{d Q^2}=-\frac{36 \alpha}{(9+12 \alpha Q)^{3 / 2}}
\end{equation}
\begin{equation}
\frac{d^2{S_1}}{d Q d \alpha}=-\frac{36 Q}{(9+12 \alpha Q)^{3 / 2}}.
\end{equation}

We can calculate the scalar curvature of the thermodynamic geometry, which is expressed as:
\begin{equation}
R({S_1})=\frac{9+10 \alpha^2 Q^2+3 \sqrt{9+12 \alpha Q}+2 \alpha Q(9+2 \sqrt{9+12 \alpha Q})}{2(3+4 \alpha Q)(3+2 \alpha Q+\sqrt{9+12 \alpha Q})}.
\end{equation}

When ${S_2}=r_{+}^2=(3 \alpha-\alpha \sqrt{9+12 \alpha Q})/(\alpha)^2$,
\begin{equation}
\frac{d^2 {S_2}}{d \alpha^2}=\frac{6\left(-3 \sqrt{3}-2 \sqrt{3} a^2 Q^2+3 \sqrt{3+4 a Q}+a Q(-6 \sqrt{3}+4 \sqrt{3+4 a Q})\right)}{a^3(3+4 a Q)^{3 / 2}},
\end{equation}
\begin{equation}
\frac{d^2{S_2}}{d Q^2}=\frac{36 \alpha}{(9+12 \alpha Q)^{3 / 2}}
\end{equation}
\begin{equation}
\frac{d^2 {S_2}}{d Q d \alpha}=\frac{36 Q}{(9+12 \alpha Q)^{3 / 2}}.
\end{equation}
The curvature scalar of the thermodynamic geometry can be calculated as:
\begin{equation}
R({S_2})=\frac{-9-10 \alpha^2 Q^2+3 \sqrt{9+12 \alpha Q}+2 \alpha Q(-9+2 \sqrt{9+12 \alpha Q})}{2(3+4 \alpha Q)(-3-2 \alpha Q+\sqrt{9+12 \alpha Q})}.
\end{equation}
The two scenarios mentioned above do not involve any phase transition.

When $\alpha>0$ and $Q=1, P=-\frac{\alpha}{4 \pi}, r_{+}$ranges from 1 to $14, \mathrm{P}$ is an algebraic number. Plot the function graph of G-T, and find the situation where the swallowtail diagram appears.(FIG.31)
\begin{figure}
\centering
\includegraphics[width=0.7\textwidth]{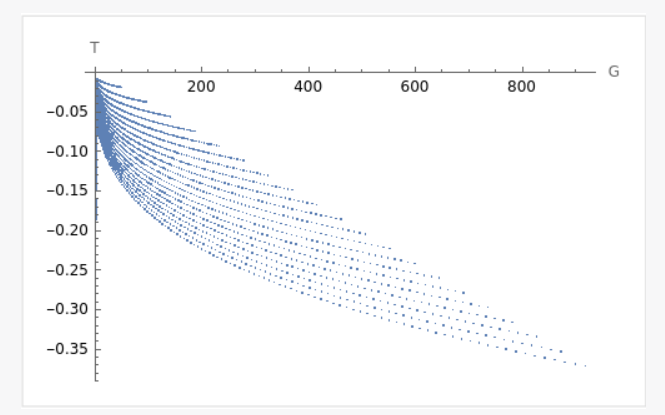}
\caption{The plot shows the relationship between the Gibbs free energy $G$ and the temperature $T$ for different values of $r_{+}$and $\alpha$. The ``swallowtail" shape in the plot represents phase transitions in the system. The exact values of $r_{+}$and $\alpha$ for which this shape appears would require a more detailed analysis.}
\label{8.11.jpg}
\end{figure}

\begin{figure}
\centering
\includegraphics[width=0.7\textwidth]{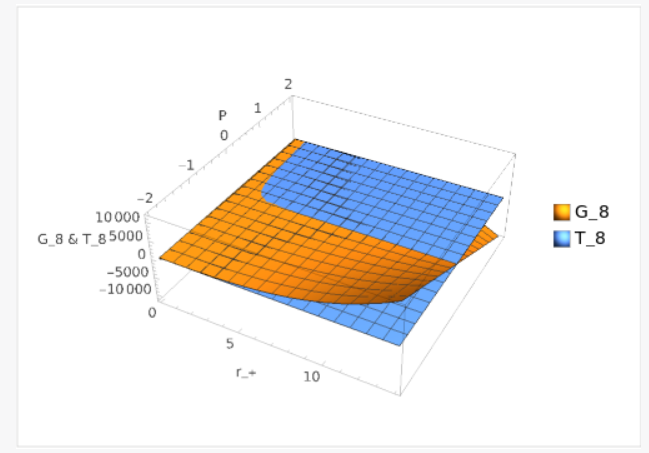}
\caption{The plot shows the relationship between the Gibbs free energy $G$ and the temperature $T$ for different values of $r_{+}$and $\alpha$. The ``swallowtail" shape in the plot represents phase transitions in the system. The exact values of $r_{+}$and $\alpha$ for which this shape appears would require a more detailed analysis.}
\label{8.111.jpg}
\end{figure}
Given(FIG.32):
\begin{equation}
\begin{aligned}
& Q=1 \\
& \alpha=\pi
\end{aligned}
\end{equation}
Let's proceed to plot $G_8$ as a function of $T_8$ and $r_{+}$.

\section{Summary and Discussion}
Significant progress has been made in inferring the possible microstructure of black holes from their macroscopic properties, involving black hole statistical thermodynamics methods. For instance, statistical thermodynamics has been employed to study phase transitions in black holes, and thermodynamic geometry has been used to infer possible phase structures and microscale interactions of black holes. Despite sharing the same four laws of thermodynamics with ordinary thermodynamic systems, black holes differ from ordinary thermodynamic systems and present many unresolved issues. In the case of ordinary thermodynamic systems such as solids or gases, the microscopic constituents are atoms or molecules, and their thermodynamic properties can be derived from statistical thermodynamics. However, for black holes, we lack the corresponding statistical thermodynamics and knowledge of their microstructure. What's even worse is that we are uncertain whether black holes have a microstructure at all. Perhaps black holes themselves are gigantic fundamental particles. In this paper, we investigate the thermodynamic geometric properties of charged black holes in the context of f(R) gravity. We find that there exists a potential barrier between the free energy G and the temperature T of charged black holes in f(R) gravity, transcending the event horizon, and under R geometry, there exists a potential well when the curvature scalar associated with entropy diverges. This provides insights into the coupling of black hole thermodynamics and quantum gravity.

We observe that when the cosmological constant has no negative power terms (i.e., the initial curvature scalar is constant), the G-T diagram shows flatness. However, when the cosmological constant has negative power terms (i.e., the initial curvature scalar is non-constant), the G-T diagram exhibits a comet-shaped structure. Although defining its negative power terms is not easy, we can still derive this conclusion.

In the background of $f(R)$ modified gravity theory, we study scalar curvature solutions with non-zero constant and discuss metric tensors satisfying the modified field equations. We investigate the thermodynamics of BHs in the absence of a cosmological constant and their local and global stability. We analyze these characteristics in different $f(R)$ models. We comment on the major differences from General Relativity and demonstrate the rich thermodynamic phenomenology characterizing this framework.In the context of the $f(R)$ modified gravity theory, we have found that certain Reissner-Nordström (RN) black holes exhibit thermodynamic behaviors similar to those of an ideal gas when the initial curvature scalar of the black hole remains constant. However, if the initial curvature scalar varies and the cosmological constant term carries a negative exponent, the Reissner-Nordström (RN) black holes may display properties resembling those of a van der Waals gas.

In this paper, we consider regular black holes in $f(R)$ gravity. Due to small perturbations near equilibrium, we summarize the expression for the modified thermodynamic entropy of this black hole. Additionally, we study the geometric thermodynamics (GTD) of the black hole and investigate the adaptability of the curvature scalar to black hole phase transitions in the geometric thermodynamics approach. Furthermore, we examine the influence of the modification parameter on the thermodynamic behavior of black holes. We provide a clear distinction between the general solutions in cases with non-negative powers and the special solutions when dealing with negative powers. Intriguingly, we note that the phase transition, bearing the resemblance to the Van der Waals gas, can materialize in charged black holes under certain conditions when subjected to f(R) gravity.

{\bf Acknowledgements:}\\
This work is partially supported by the National Natural Science Foundation of China(No. 11873025).We would like to thank Zhan-Feng Mai at Peking University for his generous help.

\end{document}